\documentclass[twocolumn,prx]{revtex4-2}
\usepackage{amsmath,amssymb,bm}
\usepackage{graphicx}

\usepackage[colorlinks=true,linkcolor=blue,urlcolor=blue,citecolor=blue]{hyperref}

\begin{document}

\title{Twisted bilayer BC$_3$: Valley interlocked anisotropic flat bands}
\author{Toshikaze Kariyado}
\affiliation{International Center for Materials Nanoarchitectonics, National Institute for Materials Science}

\begin{abstract}
Here we propose BC$_3$, a graphene derivative that has been synthesized, as a platform to realize exotic quantum phases by introducing a moir\'e pattern with mismatched stacking. In twisted bilayer BC$_3$, it is shown that a crossover from two-dimensional to quasi one-dimensional band structure takes place with the twist angle as a control parameter. This is a typical manifestation of the guiding principle in van der Waals stacked systems: the quantum interference between Bloch wave functions in adjacent layers has a striking effect on the effective interlayer tunneling. Interestingly, quasi one-dimensionalization happens in a valley dependent manner. Namely, there is interlocking between the valley index and the quasi-1D directionality, which makes BC$_3$ a plausible candidate for valleytronics devices. In addition, the strongly correlated regime of the valley interlocked quasi-1D state reduces to an interesting variant of the Kugel-Khomskii model where intertwined valley and spin degrees of freedom potentially induces exotic quantum phases. Notably, this variant of the Kugel-Khomskii model cannot be realized in conventional solids due to the three fold valley degeneracy.
\end{abstract}

\maketitle

\section{Introduction}

Nanoscale moir\'e patterns in van der Waals (vdW) heterostructures \cite{Geim:2013aa} with small interlayer mismatches currently attracts much attention as a key to manipulate electrons artificially. A typical setup to realize a tunable moir\'e pattern is a bilayer system with angle misalignment between the layers, known as a twisted bilayer system \cite{PhysRevLett.99.256802,doi:10.1021/nl902948m,Li:2010wp,PhysRevB.81.161405,PhysRevB.82.121407,Bistritzer12233,PhysRevLett.106.126802,PhysRevB.85.195458,PhysRevB.86.155449,PhysRevB.92.155409,Kim3364,Cao:2018tp,Cao:2018wy,PhysRevX.8.031087,doi:10.1126/science.aaw3780,Lu:2019ww,doi:10.1126/science.aav1910,Wong:2020vy,Zondiner:2020vo,doi:10.1126/science.aay5533,Nuckolls:2020wp,Choi:2021vi}. A moir\'e patterned bilayer system is featured by its position dependence of the relation between two layers, which induces position dependence of the effective interlayer tunneling and the electrostatic potential from the partner layers. Generically, spatial profiles of the tunneling and electrostatic potential affect the motion of electrons, allowing us to control the electronic band structures by modulating the moir\'e pattern, say by changing the twist angle. 

Regarding spatial profiles of the interlayer tunneling, the quantum interference between the Bloch wave functions in the two layers play an essential role \cite{PhysRevResearch.1.033076}. Notably, if the focused state has finite lattice momentum in each layer, the spatial profile of the effective tunneling may not follow the crystalline symmetry of the underlying lattice due to the interference effects, which potentially leads to interesting phenomena like anisotropic band flattening. In this perspective, a multivalley system, which is featured by degeneracy between the states with different momenta, is an interesting building block for twisted bilayers. 
The valley degrees of freedom is well-defined and practically conserved if the moir\'e pattern is smooth enough for electrons to suppress the intervalley scattering with large momentum transfer. If we can manipulate an electronic structure of a multivalley twisted bilayer in a valley dependent manner, it paves a way to build a moir\'e based valleytronics \cite{Rycerz:2007vn} devices.

Amongst two-dimensional (2D) atomic layer materials, graphene derivatives are interesting and important in the vdW heterostructure framework. One example is monolayer BC$_3$, where quarter of carbon atoms in graphene are replaced by boron atoms in an arrangement depicted in Fig.~\ref{fig:lattice_and_band}(a). This structure preserves the space group symmetry of graphene, but forms $2\times 2$ supercell. This BC$_3$ monolayer sheet has been synthesized on the NbB$_2$(0001) surface \cite{TANAKA200522,YANAGISAWA20064072} and some experimental characterizations have been worked out \cite{PhysRevLett.93.177003,UENO20063518}.
It is predicted to be a 2D semiconductor, i.e., the first-principles calculation shows a band gap at the Fermi energy \cite{WENTZCOVITCH1988515,PhysRevB.50.18360,BEHZAD201737,doi:10.1021/acsomega.8b01998,doi:10.1063/1.5122678,PhysRevApplied.14.014073,D0CP04219F}. Notably, the conduction bottom is at the M-point, showing anisotropic electron-like parabolic band dispersion. There are three symmetrically related M-points in the Brillouin zone due to the space group symmetry of this system, which means that monolayer BC$_3$ is a three valley system.
Having multiple valleys on high symmetric points in the Brillouin zone makes BC$_3$ a promising candidate for realizing valleytronics with the vdW heterostructure framework. 

In this paper, electronic properties of twisted bilayer BC$_3$ are investigated. For this purpose, we build two types of effective models, a tight-binding and continuum model, to have clear and intuitive understanding of the results by comparison. The parameters required in the effective models are fixed within the \textit{local approximation}, which stands on the fact that the atomic structure in a slightly twisted bilayer is \textit{locally} well approximated by the one in an untwisted bilayer system with in-plane displacement. Since the first-principles calculations can be applied on the untwisted bilayer without huge computational costs, it is used to fix the parameters in the effective models [see Fig.~\ref{fig:strategy} for the strategy to derive the effective models]. It is revealed that both of the tight-binding and the continuum models consistently show valley dependent anisotropic band flattening, in which originally 2D band dispersion is squeezed into quasi one-dimensional (1D) one with valley dependent dispersive directions upon twist. This valley interlocking quasi 1D dispersion indicates valley dependent transports, which makes twisted bilayer BC$_3$ useful for valleytronics devices.
Having the valleytronics application in mind, we also discuss possible valley selection by linearly polarized light.

In addition to the band dispersion manipulation at the noninteracting level, the strongly correlated regime of the twisted bilayer BC$_3$ is also investigated. It is shown that the strong coupling limit of the three valley effective model at the filling of one electron per a moir\'e unit cell leads to a three orbital variant of the Kugel-Khomskii model, in which a characteristic valley-spin coupling gives rise to novel quantum phases with intertwined valleys and spins. Notably, a model with the same symmetry as this generalized Kugel-Khomskii model cannot be realized in a conventional solid, due to the three fold valley degeneracy at the small angle limit where the valley degrees of freedom decouple.

The paper is organized as follows. First, we introduce effective models, and explain how to fix the required parameters within the local approximation. Next, crystalline and electronic structures for monolayer and bilayer systems are shown in order. Then, discussions for the valley selection by linearly polarized light and analysis in the strongly correlated regime are given before the summary and discussion.

\section{Methods}
\subsection{Local approximation}
\begin{figure}
    \centering
    \includegraphics{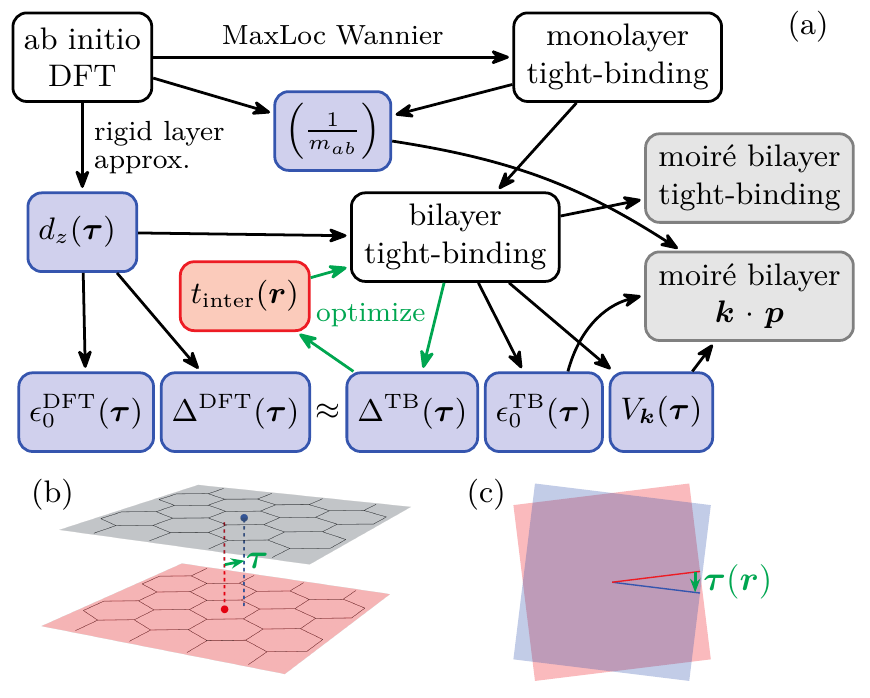}
    \caption{(a) Schematic illustration of the theoretical framework. The gray (white) boxes are for the models with (without) moir\'{e} superstructure. The blue boxes denote the quantities derived in the numerical calculation. The red box shows the empirical given function where the required parameters are chosen to optimize $\Delta^{\text{TB}}(\bm{\tau})$. (b,c) Schematic illustrations of the in-plane displacement $\bm{\tau}$ for the untwisted case (b) and the position dependent $\bm{\tau}(\bm{r})$ for the twisted case (c).}
    \label{fig:strategy}
\end{figure}
To build effective models for analyzing the electronic structure of BC$_3$, we mostly rely on the \textit{local approximation} \cite{PhysRevB.104.125427}. In the local approximation, we make full use of the fact that a crystal structure of a twisted bilayer is \textit{locally} well approximated by its untwisted counterpart for small twist angles. The untwisted bilayers can be analyzed without huge computational resources, since the unit cell size is conserved in untwisted cases.

One of the important parameters to characterize a given untwisted bilayer is an in-plane displacement $\bm{\tau}$ of the upper layer relative to the lower layer [Fig.~\ref{fig:strategy}(b)]. Then, the essence of the moir\'{e} strucutre is the position dependence of $\bm{\tau}$. In the case of a twisted bilayer, $\bm{\tau}$ depends on the position as \cite{PhysRevB.89.205414}
\begin{equation}
 \bm{\tau}(\bm{r}) = 2\sin\frac{\phi}{2}\hat{z}\times\bm{r}\label{eq:tau_r}
\end{equation}
where $\phi$ is a twist angle, $\hat{z}$ is a unit vector perpendicular to the 2D system, and $\bm{r}$ is in-plane position measured from the rotation center [see Fig.~\ref{fig:strategy}(c)]. When $\bm{\tau}(\bm{r})$ for two different positions, say $\bm{r}'$ and $\bm{r}''$, differ only by a lattice vector $n_1\bm{a}_1+n_2\bm{a}_2$ ($n_i$: integer, $\bm{a}_i$: unit vectors of the monolayer), $\bm{r}'$ and $\bm{r}''$ are identified in terms of the moir\'e pattern, giving us unit vectors for the moir\'e pattern $\bm{L}_i$ as
\begin{equation}
    \bm{L}_i = \frac{\hat{z}\times\bm{a}_i}{2\sin\frac{\phi}{2}}.
\end{equation}
Then, writing a spatial position as $\bm{r}=r_1\bm{L}_1+r_2\bm{L}_2$, and introducing $\tau_i(\bm{r})$ as $\bm{\tau}(\bm{r})=\tau_1(\bm{r})\bm{a}_1+\tau_2(\bm{r})\bm{a}_2$, Eq.~\eqref{eq:tau_r} results in \cite{PhysRevResearch.1.033076,PhysRevB.104.125427}
\begin{equation}
    \tau_i(\bm{r}_i) = -r_i.
\end{equation}
This means that scanning over the moir\'e unit cell corresponds to scanning over all possible $\bm{\tau}$ in the monolayer unit cell.

When the electronic structure of each layer is effectively approximated by a single band model, the most important effect of constructing a bilayer is interlayer tunneling, which generically lifts the band degeneracy caused by layer doubling. Specifically, $\bm{\tau}$ dependence of the splitting (and the shift of the average energy) of the target band gives important information. For untwisted bilayers, energy splitting and shift can be derived for each momentum $\bm{k}$ in the original Brillouin zone, since the unit cell size is conserved. For later use, we write these $\bm{\tau}$ dependent splitting and shift as $2\Delta_{\bm{k}}(\bm{\tau})$ and $\epsilon_{\bm{k}}(\bm{\tau})$, respectively. 

Generically, the position dependence of $\bm{\tau}$ induces the position dependence of the interlayer distance, or corrugation \cite{PhysRevB.90.155451}. In the small (but not too small) twist angle limit, this position dependence of the interlayer distance is expected to be captured by the $\bm{\tau}$ dependence of the interlayer distance in the untwisted bilayers. Taking this approximation, in the following, we first derive $\bm{\tau}$ dependent interlayer distance $d_z(\bm{\tau})$, and use that information to investigate the electronic properties. If the twist angle is too small, each layer can experiences in-plane distortions, but it is beyond our current scope.

\subsection{Effective models}
As we have noted, we are going to build two complementary effective models for BC$_3$, a tight-binding model and a $\bm{k}\cdot\bm{p}$ type continuum model. Let us start with introducing a monolayer tight-binding model, which involves $\pi$ electrons, or $p_z$-orbitals of B and C atoms. Although we have two species of the $p_z$ orbitals, from B or C, there is only one orbital per a site, allowing us to identify the site index and the orbital index. Then, the Hamiltonian for the monolayer tight-binding model is written as
\begin{equation}
 H^{\text{TB}}_{\text{mono}} = \sum_{ij}t_{ij}c^\dagger_{i}c_{j},
\end{equation}
where $c_i$ ($c^\dagger_i$) is annihilation (creation) operator for the $i$th site. Since a unit cell contains eight sites (six C atoms and two B atoms), this results in an eight band model. 

Assuming a very simple form of interlayer hoppings, we can lift this monolayer tight-binding model to a bilayer tight-binding model as
\begin{equation}
    H^{\text{TB}} = \sum_{ij\alpha}\bigl[t_{ij}c^\dagger_{i\alpha}c_{j\alpha}+t_{\text{inter}}(\bm{r}_i-\bm{r}_j)c^\dagger_{i\alpha}c_{j\bar{\alpha}}\bigl],\label{eq:def_H_TB}
\end{equation}
where $\bm{r}_i$ is the position of the $i$th site and $\alpha$ specifies the layer (upper or lower) with $\bar{\alpha}$ denoting the opposite layer of $\alpha$. Here, our simple assumption is that the interlayer hopping only depends on the relative position between the two sites. We even neglect the difference between the atom species B and C, but later it turns out to be sufficient for our purpose. Formally, Eq.~\eqref{eq:def_H_TB} can be used for both of the twisted and the untwisted bilayers, but of course the position of each site is different for the two cases, resulting in different periodicity. 

For the $\bm{k}\cdot\bm{p}$ type continuum model, we focus on the conduction bottom at M$_i$ point. For the valley M$_i$, the monolayer continuum Hamiltonian becomes
\begin{equation}
    H_{\text{mono}}^{(i)}(\bm{k}) = \epsilon_0 + \frac{\hbar^2}{2m_{\parallel}}((\bm{k}-\bm{\kappa}_{i})\cdot\bm{e}_{i}^\parallel)^2
    + \frac{\hbar^2}{2m_{\perp}}(\bm{k}\cdot\bm{e}_{i}^\perp)^2, \label{eq:def_mono_kp}
\end{equation}
where $\bm{\kappa}_i$ is the momentum for M$_i$ point and $\bm{e}_i^{\parallel,\perp}$ are unit vectors parallel and perpendicular to $\bm{\kappa}_i$. Because of the symmetry at the M-points, the relevant mass tensor is fixed by two parameters $m_\perp$ and $m_\parallel$. 

The continuum Hamiltonian for the untwisted bilayers for the M$_i$ valley can now be written as \cite{PhysRevB.95.245401}
\begin{equation}
    H^{(i)}(\bm{k},\bm{\tau}) = 
    \begin{pmatrix}
    H_{\text{mono}}^{(i)}(\bm{k}) + U_{\bm{k}}(\bm{\tau}) & V_{\bm{k}}(\bm{\tau})\\
        (V_{\bm{k}}(\bm{\tau}))^\dagger & H_{\text{mono}}^{(i)}(\bm{k}) + U_{\bm{k}}(\bm{\tau})
    \end{pmatrix},\label{eq:def_Hkp_bilayer}
\end{equation}
which is valid for $\bm{k}\sim\bm{\kappa}_i$. Here, $V_{\bm{k}}(\bm{\tau})$ represents the interlayer tunneling, while $U_{\bm{k}}(\bm{\tau})$ is contributed from the electrostatic potential from the partner layer and the multiband effects such as virtual hopping to the other bands. Both of $V_{\bm{k}}(\bm{\tau})$ and $U_{\bm{k}}(\bm{\tau})$ depend on $\bm{k}$ and $\bm{\tau}$. Since the energy eigenvalues of Eq.~\eqref{eq:def_Hkp_bilayer} are
\begin{equation}
    E^{(i)}_\pm(\bm{k},\bm{\tau}) = H^{(i)}_{\text{mono}}(\bm{k},\bm{\tau})+U_{\bm{k}}(\bm{\tau})
        \pm |V_{\bm{k}}(\bm{\tau})|,
\end{equation}
$V_{\bm{k}}(\bm{\tau})$ and $U_{\bm{k}}(\bm{\tau})$ can be related to the energy split $2\Delta_{\bm{k}}(\bm{\tau})$ and shift $\epsilon_{\bm{k}}(\bm{\tau})$ as \cite{PhysRevB.104.125427}
\begin{equation}
    \Delta_{\bm{\kappa}_i}(\bm{\tau}) = |V_{\bm{\kappa}_i}(\bm{\tau})|,\quad
    \epsilon_{\bm{\kappa}_i}(\bm{\tau}) = \epsilon_0 + U_{\bm{\kappa}_i}(\bm{\tau}).\label{eq:delta_and_epsilon}
\end{equation}

Lastly, the continuum Hamiltonian for the twisted bilayers can be written as \cite{PhysRevLett.99.256802,PhysRevB.81.161405,Bistritzer12233}
\begin{equation}
    H^{(i)} = 
    \begin{pmatrix}
    H^{(i)}_+(-i\bm{\nabla}) + U^{(i)}(\bm{r}) & V^{(i)}(\bm{r}) \\
        (V^{(i)}(\bm{r}))^\dagger & H^{(i)}_-(-i\bm{\nabla}) + U^{(i)}(\bm{r})
    \end{pmatrix},\label{eq:kp_moire}
\end{equation}
where $H^{(i)}_{\pm}$ is $H^{(i)}_{\text{mono}}$ rotated by $\pm\phi/2$. In order to take account of the spatial dependence of $U^{(i)}(\bm{r})$ and $V^{(i)}(\bm{r})$, $\bm{k}$ in $H^{(i)}_{\text{mono}}$ is replaced by $-i\bm{\nabla}$. The remaining task is to fix $U^{(i)}(\bm{r})$ and $V^{(i)}(\bm{r})$. 
In the twisted bilayer, the large scale moir\'e pattern makes the Brillouin zone folded into small moir\'e Brillouin zone, and if we further limit ourselves to small energy range, it typically becomes that the target state is mostly contributed from a small region around a certain momentum in the original Brillouin zone. In our case, Eq.~\eqref{eq:def_mono_kp} suggests that $\bm{\kappa}_i$ is regarded as the ``certain momentum''. Then, a possible approximation is to set 
\begin{equation}
    U^{(i)}(\bm{r}) = U_{\bm{\kappa}_i}(\bm{\tau}(\bm{r})),\quad
     V^{(i)}(\bm{r}) = V_{\bm{\kappa}_i}(\bm{\tau}(\bm{r})).\label{eq:UV_in_local_approx}
\end{equation}
Note that the position dependence is included through the spatial dependence of $\bm{\tau}$ under the twist. 

Combining Eqs.~\eqref{eq:delta_and_epsilon} and \eqref{eq:UV_in_local_approx}, we can fix $U^{(i)}(\bm{r})$ by $\epsilon_{\bm{\kappa}_i}(\bm{\tau})$. On the other hand, $\Delta_{\bm{\kappa}_i}(\bm{\tau})$ can fix $V_{\bm{\kappa}_i}(\bm{\tau})$ only up to the phase ambiguity, and therefore, in practice, we derive an approximate $V_{\bm{\kappa}_i}(\bm{\tau})$ using the tight-binding model. Specifically, using the Bloch wave function $\psi_{\bm{\kappa}_i}(\bm{r})$ for the target band obtained in the tight-binding model, we use \cite{PhysRevB.86.155449,PhysRevB.87.205404,PhysRevB.89.205414,PhysRevB.95.115429,PhysRevX.8.031087,PhysRevB.100.035101}
\begin{equation}
    V_{\bm{\kappa}_i}(\bm{\tau}) = e^{-i\bm{\kappa}_i\cdot\bm{\tau}}\sum_{ij}\psi^*_{\bm{\kappa}_i}(\bm{r}_i)t_{\text{inter}}(\bm{r}_i-\bm{r}_j-\bm{\tau})\psi_{\bm{\kappa}_i}(\bm{r}_j),\label{eq:Vktau_from_TB}
\end{equation}
which corresponds to the interlayer tunneling calculated at the lowest order in $t_{\text{inter}}(\bm{r})$. 

\subsection{Theoretical and numerical scheme}
Our scheme to fix the required parameters in each of the effective models is summarized in Fig.~\ref{fig:strategy}. We start with applying the first-principles density functional theory (DFT) on monolayer BC$_3$ (``ab initio DFT'' in Fig.~\ref{fig:strategy}). We first perform the lattice relaxation to obtain the crystalline parameters for the monolayer. Then, we derive the maximally localized Wennier functions to obtain the monolayer tight-binding model (``monolayer tight-binding'' in Fig.~\ref{fig:strategy}). The effective mass used in the continuum model can be derived from the band structure obtained either in DFT or the tight-binding model. 

For the bilayers, we start with deriving $d_z(\bm{\tau})$ by DFT.
Using the obtained $d_z(\bm{\tau})$ as an input, we calculate $\epsilon_{\bm{\kappa}_i}(\bm{\tau})$ and $\Delta_{\bm{\kappa}_i}(\bm{\tau})$ by DFT ($\epsilon_0^{\text{DFT}}(\bm{\tau})$ and $\Delta^{\text{DFT}}(\bm{\tau})$ in Fig.~\ref{fig:strategy}) as references for the following analysis. A tight-binding model for the untwisted bilayer (``bilayer tight-binding'' in Fig.~\ref{fig:strategy}) requires reasonable $t_{\text{inter}}(\bm{r})$, and the reasonable choice is made by calculating $\epsilon_{\bm{\kappa}_i}(\bm{\tau})$ and $\Delta_{\bm{\kappa}_i}(\bm{\tau})$ within the tight-binding model ($\epsilon_0^{\text{TB}}(\bm{\tau})$ and $\Delta^{\text{TB}}(\bm{\tau})$ in Fig.~\ref{fig:strategy}), and adjusting $t_{\text{inter}}(\bm{r})$ to have consistency between $\Delta^{\text{TB}}(\bm{\tau})$ and $\Delta^{\text{DFT}}(\bm{\tau})$. The adjusted $t_{\text{inter}}(\bm{r})$ is transported to the tight-binding model for twisted bilayers (``moir\'e bilayer tight-binding'' in Fig.~\ref{fig:strategy}) through Eq.~\eqref{eq:def_H_TB}. Or, the adjusted $t_{\text{inter}}(\bm{r})$ is plugged into Eq.~\eqref{eq:Vktau_from_TB} to have $V_{\bm{\kappa}_i}(\bm{\tau})$ required in the $\bm{k}\cdot\bm{p}$ type continuum model for the twisted bilayer (``moir\'e bilayer $\bm{k}\cdot\bm{p}$''). We can crosscheck the results under moir\'e structures by comparing the moir\'e bilayer tight-binding model and the moire bilayer $\bm{k}\cdot\bm{p}$ model. 

Before moving on to the results, we make technical notes on the numerical calculations in this paper. The first-principles calculations are performed mostly using the Quantum Espresso package \cite{Giannozzi_2009,Giannozzi_2017} with the required pseudopotentials from pslibrary \cite{pslibrary,DALCORSO2014337}. Only in one occasion, where we compare the band structures with moir\'e patterns obtained in the effective models and in the first-principles calculations, we employ the OpenMX package \cite{openmx,PhysRevB.67.155108,PhysRevB.69.195113,PhysRevB.72.045121} in which the order-$N$ calculations are implemented. Whenever the crystal structures are relevant (the lattice relaxation and the derivation of $d_z(\bm{\tau})$), rev-vdW-DF2 is used as a functional \cite{PhysRevB.76.125112,PhysRevB.89.121103} to take into account the van der Waals interaction. In other cases, including the cases where the OpenMX package is employed, we use the PBE-GGA functional \cite{PhysRevLett.77.3865} for simplicity. In the crystal structure optimization, the atomic coordinates in the unit cell and the in-plane lattice constant are relaxed until the computed forces on each atom and the in-plane cell pressure go below $1.0\times 10^{-5}$ Ry/Bohr and $5.0\times 10^{-2}$ kbar, respectively.

\section{Results}
\subsection{Monolayer}
\begin{figure}[tbp]
    \centering
    \includegraphics{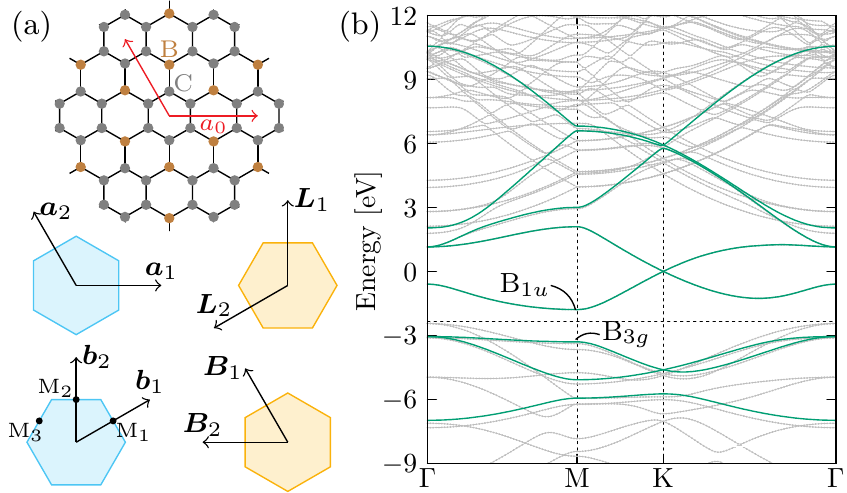}
    \caption{(a) Schematic picture for the crystal structure of monolayer BC$_3$ and the associated unit cells and Brillouin zones with (orange) and without (cyan) moir\'{e} superlattice. The monolayer lattice constant $a_0$ is 5.167 \AA. The C-C and B-C distances are 1.420 {\AA} and 1.563 \AA, respectively. The unit cells and Brillouin zones are shown to see the angle and directions, i.e., the actual sizes are very different for with and without moir\'{e} superstructures. (b) Band structure obtained with the first-principles method (gray lines), and band structure for the monolayer tight-binding model derived by the maximally localized Wannier function method (green lines).} 
    \label{fig:lattice_and_band}
\end{figure}

Let us start with the crystal structure analysis. The crystal structure of BC$_3$ [see \ref{fig:lattice_and_band}(a)] is fixed by three parameters: the in-plane lattice constant $a_0$, the B-C distance $d_{\text{B-C}}$, and the C-C distance $d_{\text{C-C}}$. They are respectively obtained as $a_0=5.167$ \AA, $d_{\text{B-C}}=1.420$ \AA, and $d_{\text{C-C}}=1.563$ \AA, consistent with the previous experimental \cite{TANAKA200522} and theoretical studies \cite{PhysRevB.37.3134,PhysRevB.50.18360,doi:10.1063/1.3681899,C3CP54838D,BEHZAD201737,doi:10.1063/1.5122678,PhysRevApplied.14.014073,D0CP04219F}.

The electronic band structure obtained by DFT with the above crystal parameters is shown as the gray lines in Fig.~\ref{fig:lattice_and_band}(b). There is a band gap at the Fermi energy (indicated as the horizontal dotted line), and the conduction bottom, on which we are going to focus, is located at the M-point showing the electron-like parabolic dispersion. At the K-point, we find a Dirac cone above the Fermi energy inherited from the pristine graphene. All of these features are consistent with the previous theoretical studies \cite{WENTZCOVITCH1988515,PhysRevB.50.18360,BEHZAD201737}. (Note that Heyd-Scuseria-Ernzerhof functional \cite{doi:10.1021/acsomega.8b01998,doi:10.1063/1.5122678,D0CP04219F} or the one-shot GW approximation \cite{PhysRevApplied.14.014073} give a larger band gap, but still the band shapes are similar to the PBE-GGA results.) As a technical note, we set the zero of energy at the energy of the Dirac cone center for the monolayer throughout the paper. 

Next, we derive a tight-binding model for the states near the Fermi energy using the Wannier90 package \cite{Pizzi2020}. Specifically, we build a tight-binding model for $\pi$ electrons, and for this purpose, we put eight $p_z$ like orbitals on the six C sites and the two B sites in the unit cell as a initial guess for the Wannier functions. We set the inner energy window, which selects an energy range where the DFT band structure is faithfully reproduced by the  tight-binding model, from $-2.07$ eV to 1.63 eV. The band structure of the obtained monolayer tight-binding model is written by the green lines in Fig.~\ref{fig:lattice_and_band}(b). It perfectly reproduces the DFT band structure in the inner energy window, including the band structure around the conduction bottom at the M-point, which plays the most important role in the following. 

From the calculated band structure, the effective mass required for the continuum models can also be derived: $m_\parallel$ and $m_\perp$ are obtained by estimating the curvature of the band at the conduction bottom at the M-point. The results are $m_\parallel=1.36 m_e$ and $m_\perp=0.15 m_e$, where $m_e$ is the bare electron mass, which is consistent with the previous work \cite{PhysRevApplied.14.014073}.

\subsection{Untwisted bilayer}
\begin{figure}[tbp]
    \centering
    \includegraphics{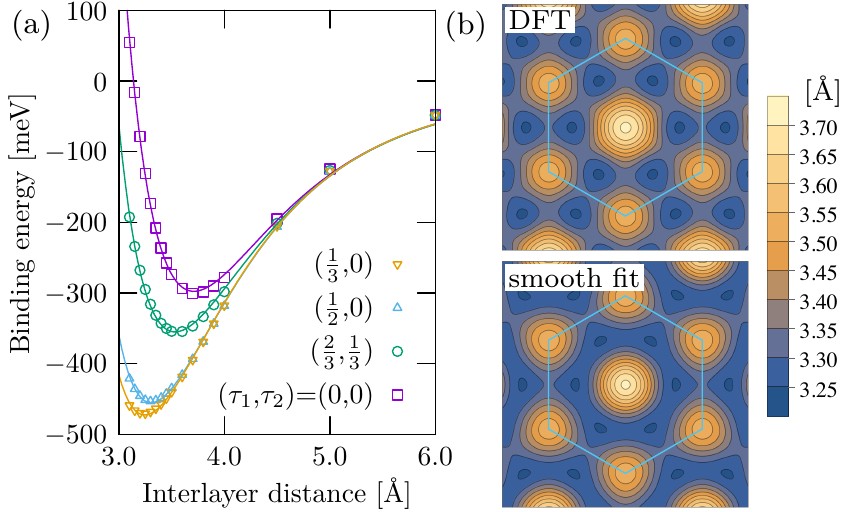}
    \caption{(a) Interlayer distance dependence of the binding energy for selected $\bm{\tau}=\tau_1\bm{a}_1+\tau_2\bm{a}_2$. The optimized layer distance for each $\bm{\tau}$ $d_z(\bm{\tau})$ is fixed by the binding energy minimum. (b) $d_z(\bm{\tau})$ obtained in the first-principles calculation (DFT) and corresponding smooth fit. The cyan hexagon denotes the unit cell.}
    \label{fig:dz}
\end{figure}

To build models for bilayers, we start with deriving the $\bm{\tau}$ dependence of the interlayer distance. For this purpose, we employ the rigid layer approximation where the interlayer distance for each $\bm{\tau}$ is determined by minimizing the total energy while the crystalline parameters for each layer are frozen. Note that the full lattice relaxation does not fit our purpose, since it automatically chooses stable (or metastable) $\bm{\tau}$, preventing the scan over the all possible $\bm{\tau}$. In practice, we set the period in $z$-direction to 40{\AA} and use the total energy for layer spacing $d_z=20${\AA} at $\bm{\tau}=0$ as a reference energy to extract binding energy. (Note that $\bm{\tau}$ dependence is negligible at $d_z=20$\AA.) Then, we plot the total energy as a function of the layer spacing for each $\bm{\tau}$, and find minima by fitting to \cite{Hsing_2014}
\begin{equation}
    E_{\text{total}}(d_z) = \alpha\exp(-\beta(d_z-d_{z0}))-\gamma(d_{z0}/d_z)^{4.5}. \label{eq:E_tot_fitting}
\end{equation}
The fitting results for the selected $\bm{\tau}$ are shown in Fig.~\ref{fig:dz}(a), and the contour plot of the obtained $d_z(\bm{\tau})$ is shown in Fig.~\ref{fig:dz}(b) with the label ``DFT''. 
The exponent of the last term in Eq.~\eqref{eq:E_tot_fitting} could be adjusted to have better fittings, but Fig.~\ref{fig:dz}(a) suggests that the long tail behavior has a limited effect on the determination of the energy minimum. 
To make the result in Fig.~\ref{fig:dz}(b) transportable to the following analysis, we approximate $d_z(\bm{\tau})$ by a simple function
\begin{multline}
    d_z(\bm{\tau}) = c_1+c_2(\cos\tau_1+\cos\tau_2+\cos(\tau_1-\tau_2))\\
    + c_3(\cos(\tau_1+\tau_2)+\cos(\tau_1-2\tau_2)+\cos(2\tau_1-\tau_2))\\
    + c_4(\cos(2\tau_1)+\cos(2\tau_2)+\cos(2\tau_1-2\tau_2))
\end{multline}
with \{$c_1$,$c_2$,$c_3$,$c_4$\}=\{3.361\AA, 0.025\AA, 0.077\AA, 0.015\AA\} whose contour plot is shown in Fig.~\ref{fig:dz}(b) with the label ``smooth fit''. We can see that this function nicely compare with the DFT result. 

\begin{figure}[tbp]
    \centering
    \includegraphics{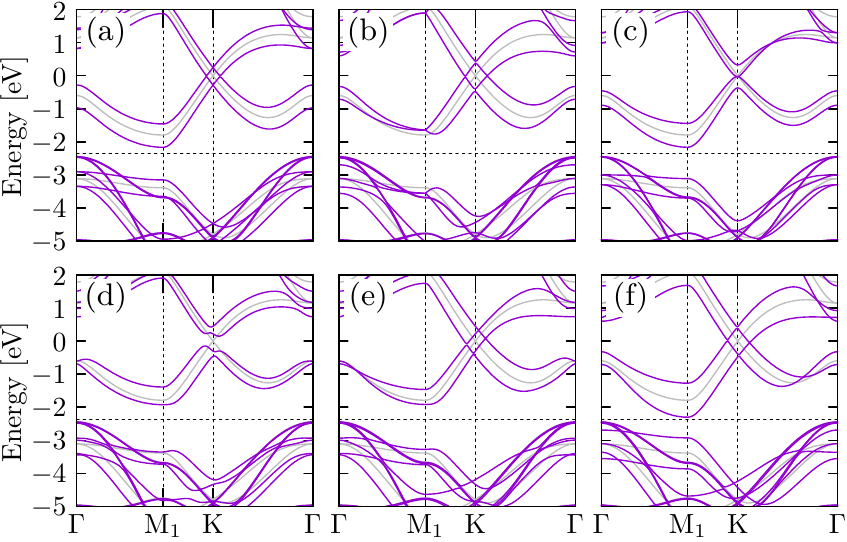}
    \caption{Band structures for untwisted bilayers with selected $\bm{\tau}$ obtained by the first-principles method. The horizontal dashed line represents the original Fermi level. (a) to (f) correspond to $(\tau_1,\tau_2)=(0,0)$, $(1/2,0)$, $(2/3,1/3)$, $(1/3,1/3)$, $(0,1/3)$, and $(0,1/2)$, respectively.}
    \label{fig:deg_lift}
\end{figure}

\begin{figure}
    \centering
    \includegraphics{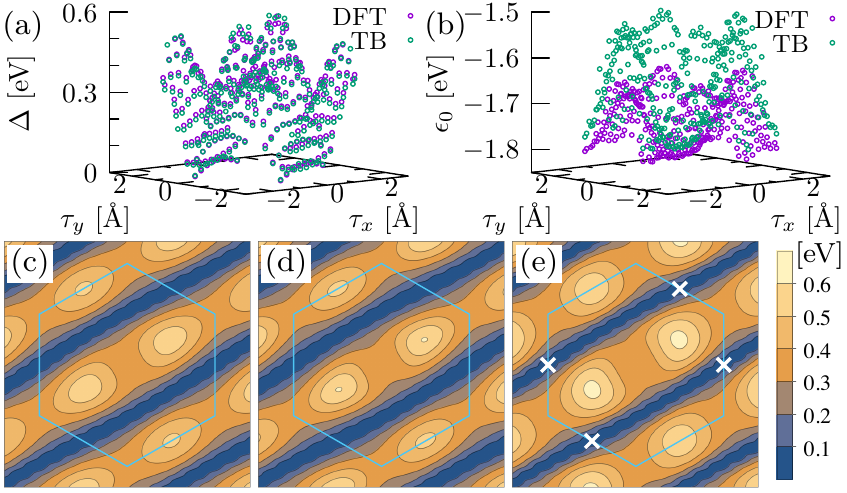}
    \caption{(a,b) Interlayer coupling induced (a) gap $\Delta(\bm{\tau})$ and (b) mean energy shift $\epsilon_0(\bm{\tau})$. The results obtained by the first-principles method (DFT) and by the tight-binding model with empirical interlayer hopping (TB) are compared. We choose the parameters in the empirical interlayer hopping to match $\Delta^{\text{DFT}}(\bm{\tau})$ and $\Delta^{\text{TB}}(\bm{\tau})$, and (a) shows that the choice is reasonable. $\epsilon_0^{\text{DFT}}(\bm{\tau})$ and $\epsilon_0^{\text{TB}}(\bm{\tau})$ are similar in shapes, but with a bit different energy scale. (c-e) Contour plots for (c) $\Delta^{\text{DFT}}(\bm{\tau})$, (d) $\Delta^{\text{TB}}(\bm{\tau})$, and (e) $|V_{\bm{\kappa}_1}(\bm{\tau})|$.} 
    \label{fig:delta}
\end{figure}

Using $d_z(\bm{\tau})$ as the interlayer distance for each $\bm{\tau}$, we inspect the $\bm{\tau}$ dependence of the electronic band structure focusing on how the interlayer tunneling modifies the band structure. Figure~\ref{fig:deg_lift} shows the band structures of monolayers (the gray lines) and bilayers (the purple lines) for the selected $\bm{\tau}$ obtained within DFT. (For some $\bm{\tau}$, the bilayer band structures can be found in the literature \cite{BEHZAD201737,PhysRevApplied.14.014073}.) The bands are split and shifted in the bilayers with $\bm{\tau}$ dependence. In our analysis, the $\bm{\tau}$ dependence of the split and the shift of the conduction bottom at the M-point is important. The split is extracted as $\Delta^{\text{DFT}}_{\bm{\kappa}_i}(\bm{\tau})=(E_2(\bm{\kappa}_i,\bm{\tau})-E_1(\bm{\kappa}_i,\bm{\tau}))/2$ while the shift is extracted as $\epsilon^{\text{DFT}}_{\bm{\kappa}_i}(\bm{\tau})=(E_1(\bm{\kappa}_i,\bm{\tau})+E_2(\bm{\kappa}_i,\bm{\tau}))/2$, where $E_1(\bm{\kappa}_i,\bm{\tau})$ and $E_2(\bm{\kappa}_i,\bm{\tau})$ are the energy of the lowest and the second lowest conduction band at the M$_i$-point, respectively. 
$\Delta^{\text{DFT}}_{\bm{\kappa}_1}(\bm{\tau})$ and $\epsilon^{\text{DFT}}_{\bm{\kappa}_1}(\bm{\tau})$ for the M$_1$ point are plotted in Figs.~\ref{fig:delta}(a) and \ref{fig:delta}(b) by the purple dots. For 
$\Delta^{\text{DFT}}_{\bm{\kappa}_1}(\bm{\tau})$, we also show its contour plot on the $\bm{\tau}$ space in Fig.~\ref{fig:delta}(c). Because of the interference between the Bloch wave functions in the upper and the lower layers, there appears characteristic quasi one-dimensional dip structure in $\Delta^{\text{DFT}}_{\bm{\kappa}_1}(\bm{\tau})$. The direction of the quasi-1D dip depends on the valley, i.e., $\Delta^{\text{DFT}}_{\bm{\kappa}_i}(\bm{\tau})$ for the M$_2$ (M$_3$) point is obtained by rotating $\Delta^{\text{DFT}}_{\bm{\kappa}_1}(\bm{\tau})$ in Fig.~\ref{fig:delta}(c) by 60$^\circ$ (120$^\circ$). This valley dependent quasi-1D structure plays a crucial role in the later discussions.

For the bilayer tight-binding model, the intralayer hopping $t_{ij}$ has already derived, and the remaining task is to determine the interlayer hopping $t_{\text{inter}}(\bm{r})$. In this study, a simple form
\begin{equation}
     t_{\text{inter}}(\bm{r})=h_0\exp\Bigl(-\frac{2d_0(r_z-d_0)}{r_0^2}\Bigr)\exp\Bigl(-\frac{r_x^2+r_y^2}{r_0^2}\Bigr),\label{eq:t_inter}
\end{equation}
is assumed where $d_0=d_z(\bm{0})$. We set $r_0$ and $h_0$ as 2.0 {\AA} and 0.30 eV, respectively.
Just as in the case of DFT, we derive the split $\Delta^{\text{TB}}_{\bm{\kappa}_i}(\bm{\tau})$ and the shift $\epsilon^{\text{TB}}_{\bm{\kappa}_i}(\bm{\tau})$ by the energies of the lowest and the second lowest conduction band at the M$_i$-point in the tight-binding model. $\Delta^{\text{TB}}_{\bm{\kappa}_1}(\bm{\tau})$ and $\epsilon^{\text{TB}}_{\bm{\kappa}_1}(\bm{\tau})$ for the M$_1$ point are plotted in Figs.~\ref{fig:delta}(a) and \ref{fig:delta}(b) by the green dots. Also, Fig.~\ref{fig:delta}(d) shows the contour plot of  $\Delta^{\text{TB}}_{\bm{\kappa}_1}(\bm{\tau})$. As Figs.~\ref{fig:delta}(a)-\ref{fig:delta}(d) show, $\Delta^{\text{TB}}_{\bm{\kappa}_i}(\bm{\tau})$ well compares with $\Delta^{\text{DFT}}_{\bm{\kappa}_i}(\bm{\tau})$, despite the simple form of $t_{\text{inter}}(\bm{r})$.
Since the parameters are chosen to have a good match between $\Delta^{\text{TB}}_{\bm{\kappa}_i}(\bm{\tau})$ and $\Delta^{\text{DFT}}_{\bm{\kappa}_i}(\bm{\tau})$, we find some quantitative difference for $\epsilon_{\bm{\kappa}_i}(\bm{\tau})$. However, the overall structures, or, the qualitative features of the $\bm{\tau}$ dependence are common to $\epsilon^{\text{TB}}_{\bm{\kappa}_i}(\bm{\tau})$ and $\epsilon^{\text{DFT}}_{\bm{\kappa}_i}(\bm{\tau})$. 

Once the interlayer hopping $t_{\text{inter}}(\bm{r})$ is fixed, it is possible to have $V_{\bm{\kappa}_i}(\bm{\tau})$ through Eq.~\eqref{eq:Vktau_from_TB}. Figure~\ref{fig:delta}(e) is the contour plot of $|V_{\bm{\kappa}_1}(\bm{\tau})|$ in the $\bm{\tau}$ space, which shows the satisfactory match to Fig.~\ref{fig:delta}(d). The reason for the satisfactory but imperfect matching is that Eq.~\eqref{eq:Vktau_from_TB} is derived at the lowest order in $t_{\text{inter}}(\bm{r})$. The previous theory on symmetry based constraints on $V_{\bm{k}}(\bm{\tau})$ \cite{PhysRevB.95.245401,PhysRevResearch.1.033076} guarantees that $V_{\bm{\kappa}_1}(\bm{\tau})$ vanishes at the points marked by white crosses in Fig.~\ref{fig:delta}(e).

\subsection{Twisted bilayer}
\begin{figure}[tbp]
    \centering
    \includegraphics{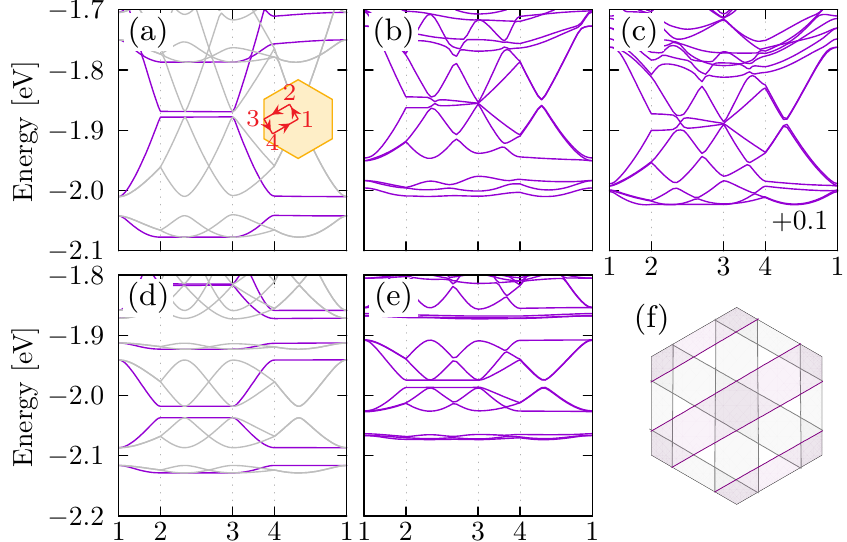}
    \caption{Band structures for twisted bilayer BC$_3$. The path on which the band structures are drawn is shown as an inset of (a). The twist angle is 7.34$^\circ$ for (a-c) and 5.09$^\circ$ for (d,e). (a) and (d) are obtained with bilayer $\bm{k}\cdot\bm{p}$ model. In (a) and (d), the purple lines are from the original M$_1$ point showing a signature of anisotropic band flattening, while the gray lines are from the original M$_2$ and M$_3$ points. (b) and (e) are obtained with moire tight-binding model, and (c) is from the large scale DFT calculation performed using OpenMX package. (f) The Fermi surfaces of the lowest energy band at half filling for the twist angle 7.34$^\circ$. As in the case for the band structures, the purple lines are contributed from the M$_1$ valley, while the gray lines are contributed from the M$_2$ and $M_3$ valleys.}
    \label{fig:band}
\end{figure}

Using $\epsilon^{\text{TB}}_{\bm{\kappa}_i}(\bm{\tau})$ and $V_{\bm{\kappa}_i}(\bm{\tau})$ obtained in the previous section, $U^{(i)}(\bm{r})$ and $V^{(i)}(\bm{r})$ in the continuum Hamiltonian Eq.~\eqref{eq:kp_moire} are fixed through Eqs.~\eqref{eq:delta_and_epsilon} and \eqref{eq:UV_in_local_approx}. The energy band structure for this continuum model can be calculated by the plane wave expansion method. The obtained band dispersions for $\phi=7.34^\circ$ and $5.09^\circ$ are shown in Figs.~\ref{fig:band}(a) and \ref{fig:band}(d), respectively. In the continuum model, each valley can be treated separately, and the purple lines are from the M$_1$ valley, while the gray lines are contributed from the M$_2$ and M$_3$ valleys. 
Focusing on the purple lines, the low energy bands are dispersive in the 1-2 or 3-4 direction, but flat in the 1-4 or 2-3 direction. That is, we find valley dependent anisotropic band flattening. Naming the dispersive direction the \textit{easy direction} (since the electrons can move in this direction easily), the valley dependent anisotropic band flattening is rephrased as the valley dependence of the easy direction. This valley dependent quasi-1D feature is also seen in the isoenergy contour (Fermi energy) shown in Fig.~\ref{fig:band}(f). Considering the fact that $V^{(i)}(\bm{r})$ looks the 90$^\circ$ rotated image of $V_{\bm{\kappa}_i}(\bm{\tau})$ due to the action of $\bm{\tau}(\bm{r})$, the 1-2 and 3-4 direction corresponds to the direction in which $V^{(i)}(\bm{r})$ only weakly vary, while the 1-4 and 2-3 direction corresponds to the direction in which $V^{(i)}(\bm{r})$ strongly vary. 
This confirms the naive expectation that $V^{(i)}(\bm{r})$ act as a potential for electrons. 

Then, for each valley, the system is regarded as a set of 1D channels aligned with the nanometer scale spacing. Under the magnetic field, this in principles can host novel phases like fractional quantum Hall states discussed in the context of the coupled wire construction, depending on the interchannel interaction. 

The electronic structure of the twisted bilayers can also be analyzed using the moir\'e tight-binding model. The Hamiltonian is Eq.~\eqref{eq:def_H_TB}, with the position of the $p_z$ orbitals determined taking account of the relative angle mismatch and the corrugation $d_z(\bm{\tau}(\bm{r}))$. Figures~\ref{fig:band}(b) and \ref{fig:band}(e) show the calculated band structures for $\phi=7.34^\circ$ and $5.09^\circ$, respectively. These angles are chosen so that the periodicity of the microscopic structure exactly matches to the periodicity of the moir\'e pattern. In the moir\'e tight-binding description, the contributions from all three valleys M$_1$ to M$_3$ come at once, and is not strictly separable. Comparing Figs.~\ref{fig:band}(a) and \ref{fig:band}(b) [or Figs.~\ref{fig:band}(d) and \ref{fig:band}(e)], we can see that some degeneracies originated from the decoupled valleys in the continuum model are lifted by the intervalley coupling in the tight-binding model. However, the band structures in the continuum model and the tight-binding model share the qualitative features. We expect that the intervalley coupling is rapidly killed in the small angle limit, because it requires large momentum transfer, while the spatial dependence of the moir\'e pattern is relatively smooth. As we have noted, Eq.~\eqref{eq:Vktau_from_TB} is an approximation for the interlayer tunneling evaluated only up to the first order in $t_{\text{inter}}(\bm{r})$, and this is a possible source of the qualitative discrepancy between the continuum model and the tight-binding model other than the intervalley coupling.

In the case that the twist angle is relatively large, it is possible to apply the first-principles method on the large moir\'e unit cell. To handle the large moire unit cell, here we employ the OpenMX package instead of the Quantum Espresso package. Figure~\ref{fig:band}(c) shows the band structure for $\phi=7.34^\circ$ obtained using the same atomic coordinates as the moir\'e bilayer tight-binding counterpart. Now, we can see that the three approaches, the effective continuum model, the effective tight-binding model, and the large scale DFT give not exactly the same but consistent band structures, which is remarkable considering the simplicity of the assumed $t_{\text{inter}}(\bm{r})$. 

\subsection{Valley selection}
As the valley dependent anisotropic band flattening is promising for valleytronics applications, it is interesting to manipulate the valley population by external perturbations. Here, we show that linearly polarized light can induce valley population imbalance. 
\begin{figure}[tbp]
    \centering
    \includegraphics{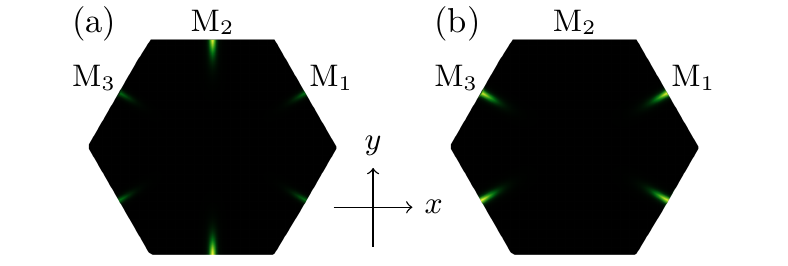}
    \caption{Momentum dependence of (a) $\Xi_{xx}(\omega,\bm{k})$ and (b) $\Xi_{yy}(\omega,\bm{k})$ at $\hbar\omega=1.5$ eV.} 
    \label{fig:optical}
\end{figure}

For simplicity, we compute the optical conductivity of the monolayer using the monolayer tight-binding model.
The real part of the optical conductivity is computed as
 \begin{equation}
  \mathrm{Re}\sigma_{\mu\mu}(\omega) = \sum_{\bm{k}}(\xi^{\mu\mu}_{\bm{k}}(\omega)-\xi^{\mu\mu}_{\bm{k}}(0))
 \end{equation}
 with
  \begin{equation}
  \xi^{\mu\mu}_{\bm{k}}(\omega) 
   = -\frac{e^2}{\omega}\sum_{\alpha\beta}
   \frac{|\tilde{v}^{\mu}_{\bm{k},\alpha\beta}|^2(f(E_{\bm{k}\alpha})-f(E_{\bm{k}\beta}))\delta}{(\omega-(E_{\bm{k}\alpha}-E_{\bm{k}\beta}))^2+\delta^2}
 \end{equation}
where $\tilde{v}^{\mu}_{\bm{k},\alpha\beta}$ is the $\mu$ component of the velocity operator in the band diagonal basis, and $E_{\bm{k}\alpha}$ is the band energy. 
To see the valley dependence, we focus on the integrand $\Xi_{\mu\mu}(\omega,\bm{k})\equiv\xi^{\mu\mu}_{\bm{k}}(\omega)-\xi^{\mu\mu}_{\bm{k}}(0)$. 
Figure~\ref{fig:optical} shows the $\bm{k}$ dependence of the obtained $\Xi_{\mu\mu}(\omega,\bm{k})$ for $\omega=1.5$ eV. Since we are now handling the monolayer model, the Brillouin zone is the original one, not the moir\'e Brillouin zone. Figures~\ref{fig:optical}(a) and \ref{fig:optical}(b) are results for the case with the light polarized in $x$ and $y$ direction, respectively. The M$_2$ valley is bright in Fig.~\ref{fig:optical}(a), while the M$_1$ and M$_3$ valleys are bright in Fig.~\ref{fig:optical}(b), which clearly indicates the valley dependent response. The symmetry analysis within DFT reveals that the irreducible representations for the highest energy valence band and the lowest energy conduction band at the M point are B$_{3g}$ and B$_{1u}$, respectively. This explains the valley dependence in Fig.~\ref{fig:optical}, since the function representations of B$_{3g}$ and B$_{1u}$ are $yz$ and $z$, respectively. 
Note that the DFT results indicates that the monolayer BC$_3$ has an indirect band gap with the valence top at the $\Gamma$-point contributed by the $\sigma$ band, while the $\sigma$ electrons are not included in our tight-binding model. Another note is that the standard DFT underestimate the gap size as we have noted, and $\omega$ to induce the valley imbalance has to be adjusted to the experimental gap size.

\subsection{Strongly correlated regime}
The purple lines in Figs.~\ref{fig:band}(a) and \ref{fig:band}(d) shows that the lowest energy band in the effective continuum model is isolated from the other band. Here, we try to have an effective description of this specific band focusing on the strongly correlated regime. Due to the three fold valley degeneracy, we will end up with a three-orbital Hubbard model, where each orbital shows quasi-1D features associated with the valley dependent easy direction. A similar model appeared in the previous study of generic twisted bilayers \cite{PhysRevResearch.1.033076}, leading to an interesting variant of the Kugel-Khomskii model in the strongly correlated limit. In the following, we derive the Kugel-Khomskii like model for twisted bilayer BC$_3$. 
\begin{figure}[tbp]
    \centering
    \includegraphics{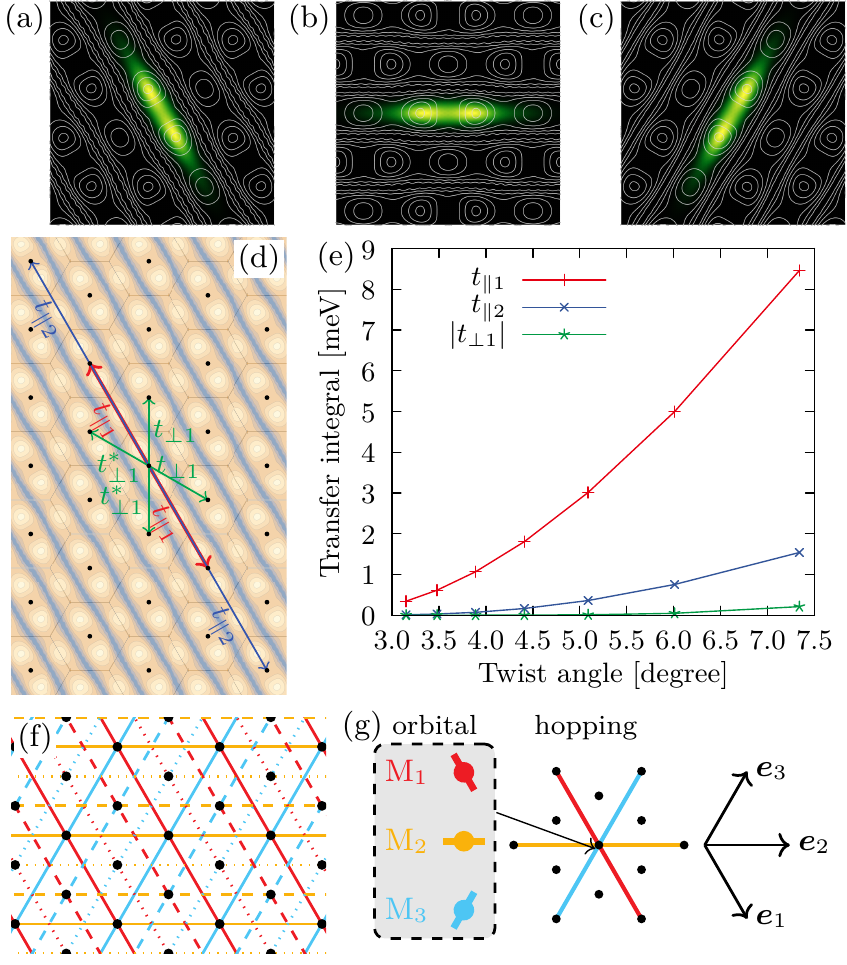}
    \caption{(a-c) Wannier functions for the lowest band in the bilayer $\bm{k}\cdot\bm{p}$ model at $\theta=5.09^\circ$. (a), (b), and (c) are for the valleys from the original M$_1$, M$_2$, and M$_3$ points, respectively. The overlaid gray lines denote the contours of $|V(\bm{r})|$. (d) Definitions of the selected hopping integrals in the tight-binding model  for the lowest band in the bilayer $\bm{k}\cdot\bm{p}$ model. The case of the M$_1$ valley is shown as an example. The background color map represents the contours of $|V(\bm{r})|$. (e) Angle dependence of the selected hopping integrals. (f) Nested network structure for the $t_{\parallel 1}$ only model. Three networks written by the solid, dashed, dotted lines are decoupled with each other in terms of the hopping. If we further truncate the offsite Hubbard interactions, we can independently treat the three networks. (g) Schematic description of the ingredients in the minimal Hubbard model for one of the three networks. There are three orbitals, originally from the three valleys, and they hop in three different directions.}
    \label{fig:hubbard}
\end{figure}

Let us start with deriving a Hubbard model, which requires basis orbitals and hopping/interaction parameters between them. For the basis orbitals and hoppings, we derive Wannier functions for the lowest energy band in the effective continuum model for each valley. This Wannier function should have a length scale of the moir\'e pattern rather than the scale of the atomic orbitals. Here, the Wannier function is derived by projecting a candidate wave function to the subspace spanned by the lowest energy band, as is done in preparing an initial guess in the well-known algorithm for maximally localized Wannier function \cite{PhysRevB.56.12847}. In our case, since the lowest energy band is well separated from the other bands, this ``initial guess'' is already well localized if we choose a proper candidate wave function, and we will use it as a Wannier function. In practice, a gaussian function whose decay length is $|\bm{L}_i|/10$ placed at the unit cell center is used as a candidate function for both of the upper and the lower layer but with the opposite sign. The obtained Wannier functions for the lowest energy bands in the M$_1$, M$_2$, and M$_3$ valleys at $\phi=5.09^\circ$ are shown in Figs.~\ref{fig:hubbard}(a)-\ref{fig:hubbard}(c), where the sum of the charge densities on the both layers is plotted. They are well localized in the region where $|V_{\bm{\kappa}_i}(\bm{\tau}(\bm{r}))|$ is large, and elongated in the corresponding easy directions.

The major hoppings for the orbital from the M$_1$ valley are illustrated in Fig.~\ref{fig:hubbard}(d). (The same thing for M$_2$ or M$_3$ can be obtained by rotating the picture appropriately.) As it should be, the major hoppings are along the easy direction. The major hoppings as functions of the twist angle are shown in Fig.~\ref{fig:hubbard}(e). The smallness of $|t_{\perp 1}|$ signals the quasi-1D feature. Also, $t_{\parallel 2}$ is significantly smaller than $t_{\parallel 1}$, and thus, keeping only $t_{\parallel 1}$ is a reasonable approximation. In this approximation, each orbital forms a 1D chain only with the nearest neighbor hopping, where the hopping direction depends on the orbitals. Note that if we turn off $t_{\perp,1}$ (and the other hoppings not along the easy direction), the system is decoupled into three nested triangular networks as is illustrated by the solid, dashed, and dotted lines in Fig.~\ref{fig:hubbard}(f).

Now, let us move on to the electron-electron interaction terms, focusing on the onsite Coulomb repulsion between the Wannier orbitals for simplicity. The onsite interaction still preserves the decoupled nature of the three nested networks. Because of the elongated shapes of the Wannier functions [Figs.~\ref{fig:hubbard}(a)-\ref{fig:hubbard}(c)], distinction between the intraorbital interaction $U$ and the interorbital interaction $U'$ is important. A rough estimation of $U$ and $U'$ is given by \cite{PhysRevX.8.031087}
  \begin{equation}
  U_{\alpha\alpha'} 
   = \frac{1}{(2\pi)^2}\int d^2\bm{q}\rho_{\alpha,\bm{q}}V_{\bm{q}}\rho_{\alpha',-\bm{q}},\label{eq:U_aa}
 \end{equation}
 with $U = U_{\alpha\alpha}$ and $U'=U_{\alpha\alpha'}$ ($\alpha\neq\alpha'$). Here, $V_{\bm{q}}$ is the Fourier components of the (screened) Coulomb interaction and $\rho_{\alpha\bm{q}}$ is the sum of the charge densities in the upper and the lower layer for the orbital $\alpha$. 

Since the three nested networks are decoupled, we pick up one out of them from now on. Then, the effective Hubbard model becomes
 \begin{multline}
  H=t_{\parallel 1}\sum_{\bm{r}\sigma}\sum_{\mu=1}^3
  c^\dagger_{\bm{r}+\bm{e}_\mu,\mu\sigma}c_{\bm{r},\mu\sigma} + \text{ h.c. }\\
  +U\sum_{\bm{r}}\sum_{\mu=1}^3 n_{\bm{r},\mu\uparrow}
  n_{\bm{r},\mu\downarrow}
  +U'\sum_{\bm{r},\sigma\sigma'}\sum_{\mu<\mu'}
  n_{\bm{r},\mu\sigma}
  n_{\bm{r},\mu'\sigma'},\label{eq:H_hubbard}
 \end{multline}
 where the orbital components and the definition of $\bm{e}_\mu$ are schematically illustrated in Fig.~\ref{fig:hubbard}(g). Expanding in terms of $t_{\parallel 1}/U$ and $t_{\parallel 1}/U'$ to the lowest order (strongly correlated limit) at $1/6$ filling, this Hamiltonian reduces to a variant of the Kugel-Khomskii model \cite{Kugel1982,PhysRevB.56.R14243,PhysRevB.80.064413}
  \begin{multline}
  H=J\sum_{\bm{r}}\sum_{\mu=1}^3
  \Bigl(\bm{S}_{\bm{r}+\bm{e}_\mu}\cdot\bm{S}_{\bm{r}}-\frac{1}{4}\Bigr)\tilde{\tau}^{(\mu)}_{\bm{r}+\bm{e}_\mu}\tilde{\tau}^{(\mu)}_{\bm{r}}\\
  -J'\sum_{\bm{r}}\sum_{\mu\neq\mu'}
  \bigl(\tilde{\tau}^{(\mu')}_{\bm{r}+\bm{e}_\mu}\tilde{\tau}^{(\mu)}_{\bm{r}}+\tilde{\tau}^{(\mu')}_{\bm{r}-\bm{e}_{\mu}}\tilde{\tau}^{(\mu)}_{\bm{r}}\bigr), \label{eq:H_KK}
 \end{multline}
with $J=4t_{\parallel 1}^2/U$, $J'=t_{\parallel 1}^2/U'$, and $\tilde{\tau}^{(\mu)}$ being a $3\times 3$ matrix in the orbital space defined as  $(\tilde{\tau}^{(\mu)})_{ij}=\delta_{ij}\delta_{i\mu}$. In going from Eq.~\eqref{eq:H_hubbard} to Eq.~\eqref{eq:H_KK}, it is important to note that the exchange is restricted by the orbital character, because an orbital only hops in its corresponding easy direction. 
 
\begin{figure}[tbp]
    \centering
    \includegraphics{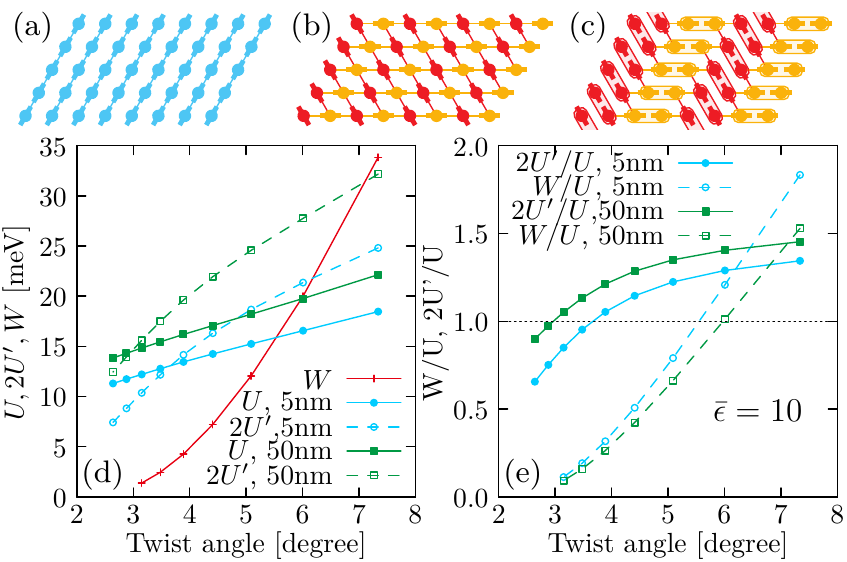}
    \caption{(a-c) Selected eigenstates of Kugel-Khomskii like model. (a) Ferro-orbital order (FOO). (b) Fully antiferro-orbital order (FAOO). (c) Dimer covering (DC). (d) Estimation of the Hubbard parameters $U$ and $U'$ by the Wannier function, compared with the band width $W=4t_{\parallel 1}$. The cyan lines are for the case with $d_{\text{gate}}=5$ nm, while the green lines are for the case with $d_{\text{gate}}=50$ nm. (e) The same as (d), but the ratios $2U'/U$ and $W/U$ are plotted.}
    \label{fig:wuv}
\end{figure}

In Eq.~\eqref{eq:H_KK}, there is no term causing the orbital flip, which means that the orbital degrees of freedom is regarded as classical, allowing us to derive some eigenstates analytically. One example is a state with the ferro-orbital order (FOO), where all the sites are in the same orbital state [Fig.~\ref{fig:wuv}(a)]. The Hamiltonian Eq.~\eqref{eq:H_KK} indicates that the Heisenberg interaction ($J$ term) works only in the easy direction of the selected orbital. Then, the FOO state is equivalent to the collection of the decoupled Heisenberg chains, and the lowest energy with FOO is obtained by the ground state energy of the Heisenberg chain as \cite{Giamarchi_1D}
\begin{equation}
    E^{\text{FOO}}_{\text{per site}}\sim -0.69 J.
\end{equation}
Note that this state is nematic, since Fig.~\ref{fig:wuv}(a) preserves the lattice translation symmetry, but breaks the rotation symmetry.
Another example is a state with the fully antiferro-orbital order (FAOO) depicted in Fig.~\ref{fig:wuv}(b). With this configuration, there is no active $J$ bonds and the energy is obtained from counting the active $J'$ bonds, resulting in
\begin{equation}
    E^{\text{FAOO}}_{\text{per site}}=-2J'.
\end{equation}
This state has macroscopic number of degeneracy. Namely, any M$_1$ and M$_2$ orbitals in the state of Fig.~\ref{fig:wuv}(b) can be replaced by M$_3$ orbitals without energy cost, as far as the introduced M$_3$ orbitals induces no ferro-orbital coupling. The last example is a state with the dimer covering (DC) [Fig.~\ref{fig:wuv}(c)]. In this configuration, all the sites are paired to dimers by the $J$ term, and the lowest energy with DC order is obtained by the energy of the spin singlet, resulting in 
\begin{equation}
    E^{\text{DC}}_{\text{per site}} = -J/2-J'.
\end{equation}
This state also has macroscopic degeneracy, which can be checked by inserting M$_3$ dimers in Fig.~\ref{fig:wuv}(c). For both of the FAOO and the DC states, the macroscopic degeneracy will be lifted by quantum fluctuation induced by, for instance, the ring exchange, and possibly lead to exotic quantum phases.

From the definitions of $J$ and $J'$, we have $E^{\text{DC}}_{\text{per site}}<E^{\text{FOO}}_{\text{per site}}$ for $U>U'$. We also have $E^{\text{DC}}_{\text{per site}}<E^{\text{FAOO}}_{\text{per site}}$ for $U<2U'$, while $E^{\text{DC}}_{\text{per site}}>E^{\text{FAOO}}_{\text{per site}}$ for $U>2U'$, i.e., there should be phase transitions controlled by $U'/U$. 

Now, let us estimate $U$ and $U'$ by Eq.~\eqref{eq:U_aa} with the special attention to the ratio between $U$ and $2U'$. For $V_{\bm{q}}$, we use
\begin{equation}
     V_{\bm{q}} = \frac{2\pi k_0 e^2}{\bar{\epsilon}q}\tanh(qd_{\text{gate}}), \label{eq:screened_Coulomb}
\end{equation}
where $k_0=1/(4\pi\epsilon_0)$ and $\bar{\epsilon}=\epsilon/\epsilon_0$. This is for the case that the Coulomb interaction is screened by metallic gates above and below the system located $d_{\text{gate}}$ off from the system \cite{PhysRevB.100.205113,PhysRevB.102.035161,PhysRevB.102.045107}. Note that $d_{\text{gate}}$ sets a length scale to the Coulomb interaction. 
In reality, we have to take care of all possible sources of screening, but here for simplicity, we stick to Eq.~\eqref{eq:screened_Coulomb} and treat $\epsilon$ as an adjustable parameter. 

The Hubbard parameter $U$ and $2U'$ computed with $\bar{\epsilon}=10$ are plotted as a function of the twist angle in Fig.~\ref{fig:wuv}(d), together with the band width estimated by $W=4|t_{\parallel 1}|$. $U$ and $U'$ decrease as the twist angle gets smaller, but the decrease is much slower than that of $W$ [see $W/U$ in Fig.~\ref{fig:wuv}(e)], leading to strongly correlated regime at the small angle. Another notable feature is that there is crossing between $U$ and $2U'$, which is captured by the point of $2U'/U=1$ in Fig.~\ref{fig:wuv}(e). As Figs.~\ref{fig:hubbard}(a)-\ref{fig:hubbard}(c) show, the Wannier functions are elongated in the corresponding easy directions. In evaluating $U'$, the two Wannier functions are misaligned, and this misalignment significantly suppresses $U'$ when the screening is strong and efficient. For fixed $d_{\text{gate}}$, the screening becomes efficient as the length scale of the Wannier functions becomes large, and this is the reason for having $U>2U'$ in the small twist angle limit. It also explains the reason that the crossing occurs at larger twist angle for smaller $d_{\text{gate}}$. 

\section{Summary and Discussions}
To summarize, twisted bilayer BC$_3$ is predicted to show valley dependent anisotropic band flattening. This is a typical (and almost ideal) realization of the known theoretical wisdom: the quantum interference between Bloch wave functions in two layers leads to interesting band engineering in twisted bilayers. 
In the application view point, the valley dependent anisotropic band flattening is promising for valleytronics devices and it is shown that the linearly polarized light induces valley imbalance. 

Also, our analysis itself gives an important perspective on theoretical/computational methodology for moir\'e systems. Namely, the current analysis serves as a working example of the cycle described in Fig.~\ref{fig:strategy}, where the effective model with moir\'e patterns is derived by the information collected without moir\'e patterns through the local approximation. The method reduces the computational cost, and helps to obtain intuitive understanding of the interlayer coupling.

In the strongly correlated regime, BC$_3$ is shown to host the three orbital Hubbard and Kugel-Khomskii like model. The three fold degeneracy, originating from the suppressed intervalley scattering, is difficult to realize in 2D, since there is no three dimensional irreducible representation in 2D space groups (without the spin-orbit coupling), which highlights the uniqueness of BC$_3$. Indeed, some interesting eigenstates are derived in our Kugel-Khomskii like model. The state in Fig.~\ref{fig:wuv}(a) preserves the lattice translation symmetry, but breaks the rotational symmetry, i.e., it is a nematic phase, while the states in Figs.~\ref{fig:wuv}(b) and \ref{fig:wuv}(c) have macroscopic degeneracy that potentially leads to novel quantum phases. In a generic perspective, it is worth to note that the twist angle can be used to tune the ratio between the intra and interorbital Hubbard parameters ($U'/U$), not only the ratio between the kinetic and the interaction energy ($U/W$). This opens up a rich avenue to explore interesting quantum phases in multivalley moir\'e systems.

Recently, there is an experimental realization of a quasi-1D state in twisted bilayer tungsten ditelluride \cite{wang2022-luttinger}. One unique feature of BC$_3$ in comparison with that former example is that the role of the quantum interference of the Bloch wave functions is evident in BC$_3$. Another unique feature of BC$_3$ is, of course, that it hosts the aforementioned generalized Kugel-Khomskii model in the strongly correlated regime. Another known candidate of the quasi-1D state realized with moir\'e pattern is twisted bilayer GeSe \cite{kennes2020one-b}, but its origin of the quasi-1D feature is different from the one in BC$_3$ \cite{PhysRevB.104.125427}.

We would like to close the paper by a comment on the experimental realization. The monolayer BC$_3$ itself has already been synthesized \cite{TANAKA200522}. Yet, it can be a big challenge to apply van der Waals stacking technique on this specific system. However, witnessing the rapid advances in the field, it will be realized in the near future.

\begin{acknowledgements}
The author thank A. Vishwanath for useful comments and discussions. This work was supported by JSPS KAKENHI Grant Number JP20K03844. Part of the computations in this work has been done using the facilities of the Supercomputer Center, the Institute for Solid State Physics, the University of Tokyo. 
\end{acknowledgements}


\begin{thebibliography}{69}%
\makeatletter
\providecommand \@ifxundefined [1]{%
 \@ifx{#1\undefined}
}%
\providecommand \@ifnum [1]{%
 \ifnum #1\expandafter \@firstoftwo
 \else \expandafter \@secondoftwo
 \fi
}%
\providecommand \@ifx [1]{%
 \ifx #1\expandafter \@firstoftwo
 \else \expandafter \@secondoftwo
 \fi
}%
\providecommand \natexlab [1]{#1}%
\providecommand \enquote  [1]{``#1''}%
\providecommand \bibnamefont  [1]{#1}%
\providecommand \bibfnamefont [1]{#1}%
\providecommand \citenamefont [1]{#1}%
\providecommand \href@noop [0]{\@secondoftwo}%
\providecommand \href [0]{\begingroup \@sanitize@url \@href}%
\providecommand \@href[1]{\@@startlink{#1}\@@href}%
\providecommand \@@href[1]{\endgroup#1\@@endlink}%
\providecommand \@sanitize@url [0]{\catcode `\\12\catcode `\$12\catcode
  `\&12\catcode `\#12\catcode `\^12\catcode `\_12\catcode `\%12\relax}%
\providecommand \@@startlink[1]{}%
\providecommand \@@endlink[0]{}%
\providecommand \url  [0]{\begingroup\@sanitize@url \@url }%
\providecommand \@url [1]{\endgroup\@href {#1}{\urlprefix }}%
\providecommand \urlprefix  [0]{URL }%
\providecommand \Eprint [0]{\href }%
\providecommand \doibase [0]{https://doi.org/}%
\providecommand \selectlanguage [0]{\@gobble}%
\providecommand \bibinfo  [0]{\@secondoftwo}%
\providecommand \bibfield  [0]{\@secondoftwo}%
\providecommand \translation [1]{[#1]}%
\providecommand \BibitemOpen [0]{}%
\providecommand \bibitemStop [0]{}%
\providecommand \bibitemNoStop [0]{.\EOS\space}%
\providecommand \EOS [0]{\spacefactor3000\relax}%
\providecommand \BibitemShut  [1]{\csname bibitem#1\endcsname}%
\let\auto@bib@innerbib\@empty
\bibitem [{\citenamefont {Geim}\ and\ \citenamefont
  {Grigorieva}(2013)}]{Geim:2013aa}%
  \BibitemOpen
  \bibfield  {author} {\bibinfo {author} {\bibfnamefont {A.~K.}\ \bibnamefont
  {Geim}}\ and\ \bibinfo {author} {\bibfnamefont {I.~V.}\ \bibnamefont
  {Grigorieva}},\ }\bibfield  {title} {\bibinfo {title} {Van der {Waals}
  heterostructures},\ }\href {https://doi.org/10.1038/nature12385} {\bibfield
  {journal} {\bibinfo  {journal} {Nature}\ }\textbf {\bibinfo {volume} {499}},\
  \bibinfo {pages} {419 EP } (\bibinfo {year} {2013})}\BibitemShut {NoStop}%
\bibitem [{\citenamefont {Lopes~dos Santos}\ \emph {et~al.}(2007)\citenamefont
  {Lopes~dos Santos}, \citenamefont {Peres},\ and\ \citenamefont
  {Castro~Neto}}]{PhysRevLett.99.256802}%
  \BibitemOpen
  \bibfield  {author} {\bibinfo {author} {\bibfnamefont {J.~M.~B.}\
  \bibnamefont {Lopes~dos Santos}}, \bibinfo {author} {\bibfnamefont
  {N.~M.~R.}\ \bibnamefont {Peres}},\ and\ \bibinfo {author} {\bibfnamefont
  {A.~H.}\ \bibnamefont {Castro~Neto}},\ }\bibfield  {title} {\bibinfo {title}
  {Graphene bilayer with a twist: Electronic structure},\ }\href
  {https://doi.org/10.1103/PhysRevLett.99.256802} {\bibfield  {journal}
  {\bibinfo  {journal} {Phys. Rev. Lett.}\ }\textbf {\bibinfo {volume} {99}},\
  \bibinfo {pages} {256802} (\bibinfo {year} {2007})}\BibitemShut {NoStop}%
\bibitem [{\citenamefont {Trambly~de Laissardi{\`e}re}\ \emph
  {et~al.}(2010)\citenamefont {Trambly~de Laissardi{\`e}re}, \citenamefont
  {Mayou},\ and\ \citenamefont {Magaud}}]{doi:10.1021/nl902948m}%
  \BibitemOpen
  \bibfield  {author} {\bibinfo {author} {\bibfnamefont {G.}~\bibnamefont
  {Trambly~de Laissardi{\`e}re}}, \bibinfo {author} {\bibfnamefont
  {D.}~\bibnamefont {Mayou}},\ and\ \bibinfo {author} {\bibfnamefont
  {L.}~\bibnamefont {Magaud}},\ }\bibfield  {title} {\bibinfo {title}
  {Localization of {Dirac} electrons in rotated graphene bilayers},\ }\href
  {https://doi.org/10.1021/nl902948m} {\bibfield  {journal} {\bibinfo
  {journal} {Nano Lett.}\ }\textbf {\bibinfo {volume} {10}},\ \bibinfo {pages}
  {804} (\bibinfo {year} {2010})}\BibitemShut {NoStop}%
\bibitem [{\citenamefont {Li}\ \emph {et~al.}(2010)\citenamefont {Li},
  \citenamefont {Luican}, \citenamefont {Lopes~dos Santos}, \citenamefont
  {Castro~Neto}, \citenamefont {Reina}, \citenamefont {Kong},\ and\
  \citenamefont {Andrei}}]{Li:2010wp}%
  \BibitemOpen
  \bibfield  {author} {\bibinfo {author} {\bibfnamefont {G.}~\bibnamefont
  {Li}}, \bibinfo {author} {\bibfnamefont {A.}~\bibnamefont {Luican}}, \bibinfo
  {author} {\bibfnamefont {J.~M.~B.}\ \bibnamefont {Lopes~dos Santos}},
  \bibinfo {author} {\bibfnamefont {A.~H.}\ \bibnamefont {Castro~Neto}},
  \bibinfo {author} {\bibfnamefont {A.}~\bibnamefont {Reina}}, \bibinfo
  {author} {\bibfnamefont {J.}~\bibnamefont {Kong}},\ and\ \bibinfo {author}
  {\bibfnamefont {E.~Y.}\ \bibnamefont {Andrei}},\ }\bibfield  {title}
  {\bibinfo {title} {Observation of {Van Hove} singularities in twisted
  graphene layers},\ }\href {https://doi.org/10.1038/nphys1463} {\bibfield
  {journal} {\bibinfo  {journal} {Nat. Phys.}\ }\textbf {\bibinfo {volume}
  {6}},\ \bibinfo {pages} {109} (\bibinfo {year} {2010})}\BibitemShut {NoStop}%
\bibitem [{\citenamefont {Mele}(2010)}]{PhysRevB.81.161405}%
  \BibitemOpen
  \bibfield  {author} {\bibinfo {author} {\bibfnamefont {E.~J.}\ \bibnamefont
  {Mele}},\ }\bibfield  {title} {\bibinfo {title} {Commensuration and
  interlayer coherence in twisted bilayer graphene},\ }\href
  {https://doi.org/10.1103/PhysRevB.81.161405} {\bibfield  {journal} {\bibinfo
  {journal} {Phys. Rev. B}\ }\textbf {\bibinfo {volume} {81}},\ \bibinfo
  {pages} {161405(R)} (\bibinfo {year} {2010})}\BibitemShut {NoStop}%
\bibitem [{\citenamefont {Su\'arez~Morell}\ \emph {et~al.}(2010)\citenamefont
  {Su\'arez~Morell}, \citenamefont {Correa}, \citenamefont {Vargas},
  \citenamefont {Pacheco},\ and\ \citenamefont
  {Barticevic}}]{PhysRevB.82.121407}%
  \BibitemOpen
  \bibfield  {author} {\bibinfo {author} {\bibfnamefont {E.}~\bibnamefont
  {Su\'arez~Morell}}, \bibinfo {author} {\bibfnamefont {J.~D.}\ \bibnamefont
  {Correa}}, \bibinfo {author} {\bibfnamefont {P.}~\bibnamefont {Vargas}},
  \bibinfo {author} {\bibfnamefont {M.}~\bibnamefont {Pacheco}},\ and\ \bibinfo
  {author} {\bibfnamefont {Z.}~\bibnamefont {Barticevic}},\ }\bibfield  {title}
  {\bibinfo {title} {Flat bands in slightly twisted bilayer graphene:
  Tight-binding calculations},\ }\href
  {https://doi.org/10.1103/PhysRevB.82.121407} {\bibfield  {journal} {\bibinfo
  {journal} {Phys. Rev. B}\ }\textbf {\bibinfo {volume} {82}},\ \bibinfo
  {pages} {121407} (\bibinfo {year} {2010})}\BibitemShut {NoStop}%
\bibitem [{\citenamefont {Bistritzer}\ and\ \citenamefont
  {MacDonald}(2011)}]{Bistritzer12233}%
  \BibitemOpen
  \bibfield  {author} {\bibinfo {author} {\bibfnamefont {R.}~\bibnamefont
  {Bistritzer}}\ and\ \bibinfo {author} {\bibfnamefont {A.~H.}\ \bibnamefont
  {MacDonald}},\ }\bibfield  {title} {\bibinfo {title} {Moir{\'e} bands in
  twisted double-layer graphene},\ }\href
  {https://doi.org/10.1073/pnas.1108174108} {\bibfield  {journal} {\bibinfo
  {journal} {Proc. Natl. Acad. Sci. U.S.A.}\ }\textbf {\bibinfo {volume}
  {108}},\ \bibinfo {pages} {12233} (\bibinfo {year} {2011})}\BibitemShut
  {NoStop}%
\bibitem [{\citenamefont {Luican}\ \emph {et~al.}(2011)\citenamefont {Luican},
  \citenamefont {Li}, \citenamefont {Reina}, \citenamefont {Kong},
  \citenamefont {Nair}, \citenamefont {Novoselov}, \citenamefont {Geim},\ and\
  \citenamefont {Andrei}}]{PhysRevLett.106.126802}%
  \BibitemOpen
  \bibfield  {author} {\bibinfo {author} {\bibfnamefont {A.}~\bibnamefont
  {Luican}}, \bibinfo {author} {\bibfnamefont {G.}~\bibnamefont {Li}}, \bibinfo
  {author} {\bibfnamefont {A.}~\bibnamefont {Reina}}, \bibinfo {author}
  {\bibfnamefont {J.}~\bibnamefont {Kong}}, \bibinfo {author} {\bibfnamefont
  {R.~R.}\ \bibnamefont {Nair}}, \bibinfo {author} {\bibfnamefont {K.~S.}\
  \bibnamefont {Novoselov}}, \bibinfo {author} {\bibfnamefont {A.~K.}\
  \bibnamefont {Geim}},\ and\ \bibinfo {author} {\bibfnamefont {E.~Y.}\
  \bibnamefont {Andrei}},\ }\bibfield  {title} {\bibinfo {title} {Single-layer
  behavior and its breakdown in twisted graphene layers},\ }\href
  {https://doi.org/10.1103/PhysRevLett.106.126802} {\bibfield  {journal}
  {\bibinfo  {journal} {Phys. Rev. Lett.}\ }\textbf {\bibinfo {volume} {106}},\
  \bibinfo {pages} {126802} (\bibinfo {year} {2011})}\BibitemShut {NoStop}%
\bibitem [{\citenamefont {Moon}\ and\ \citenamefont
  {Koshino}(2012)}]{PhysRevB.85.195458}%
  \BibitemOpen
  \bibfield  {author} {\bibinfo {author} {\bibfnamefont {P.}~\bibnamefont
  {Moon}}\ and\ \bibinfo {author} {\bibfnamefont {M.}~\bibnamefont {Koshino}},\
  }\bibfield  {title} {\bibinfo {title} {Energy spectrum and quantum {Hall}
  effect in twisted bilayer graphene},\ }\href
  {https://doi.org/10.1103/PhysRevB.85.195458} {\bibfield  {journal} {\bibinfo
  {journal} {Phys. Rev. B}\ }\textbf {\bibinfo {volume} {85}},\ \bibinfo
  {pages} {195458} (\bibinfo {year} {2012})}\BibitemShut {NoStop}%
\bibitem [{\citenamefont {Lopes~dos Santos}\ \emph {et~al.}(2012)\citenamefont
  {Lopes~dos Santos}, \citenamefont {Peres},\ and\ \citenamefont
  {Castro~Neto}}]{PhysRevB.86.155449}%
  \BibitemOpen
  \bibfield  {author} {\bibinfo {author} {\bibfnamefont {J.~M.~B.}\
  \bibnamefont {Lopes~dos Santos}}, \bibinfo {author} {\bibfnamefont
  {N.~M.~R.}\ \bibnamefont {Peres}},\ and\ \bibinfo {author} {\bibfnamefont
  {A.~H.}\ \bibnamefont {Castro~Neto}},\ }\bibfield  {title} {\bibinfo {title}
  {Continuum model of the twisted graphene bilayer},\ }\href
  {https://doi.org/10.1103/PhysRevB.86.155449} {\bibfield  {journal} {\bibinfo
  {journal} {Phys. Rev. B}\ }\textbf {\bibinfo {volume} {86}},\ \bibinfo
  {pages} {155449} (\bibinfo {year} {2012})}\BibitemShut {NoStop}%
\bibitem [{\citenamefont {Wong}\ \emph {et~al.}(2015)\citenamefont {Wong},
  \citenamefont {Wang}, \citenamefont {Jung}, \citenamefont {Pezzini},
  \citenamefont {DaSilva}, \citenamefont {Tsai}, \citenamefont {Jung},
  \citenamefont {Khajeh}, \citenamefont {Kim}, \citenamefont {Lee},
  \citenamefont {Kahn}, \citenamefont {Tollabimazraehno}, \citenamefont
  {Rasool}, \citenamefont {Watanabe}, \citenamefont {Taniguchi}, \citenamefont
  {Zettl}, \citenamefont {Adam}, \citenamefont {MacDonald},\ and\ \citenamefont
  {Crommie}}]{PhysRevB.92.155409}%
  \BibitemOpen
  \bibfield  {author} {\bibinfo {author} {\bibfnamefont {D.}~\bibnamefont
  {Wong}}, \bibinfo {author} {\bibfnamefont {Y.}~\bibnamefont {Wang}}, \bibinfo
  {author} {\bibfnamefont {J.}~\bibnamefont {Jung}}, \bibinfo {author}
  {\bibfnamefont {S.}~\bibnamefont {Pezzini}}, \bibinfo {author} {\bibfnamefont
  {A.~M.}\ \bibnamefont {DaSilva}}, \bibinfo {author} {\bibfnamefont {H.-Z.}\
  \bibnamefont {Tsai}}, \bibinfo {author} {\bibfnamefont {H.~S.}\ \bibnamefont
  {Jung}}, \bibinfo {author} {\bibfnamefont {R.}~\bibnamefont {Khajeh}},
  \bibinfo {author} {\bibfnamefont {Y.}~\bibnamefont {Kim}}, \bibinfo {author}
  {\bibfnamefont {J.}~\bibnamefont {Lee}}, \bibinfo {author} {\bibfnamefont
  {S.}~\bibnamefont {Kahn}}, \bibinfo {author} {\bibfnamefont {S.}~\bibnamefont
  {Tollabimazraehno}}, \bibinfo {author} {\bibfnamefont {H.}~\bibnamefont
  {Rasool}}, \bibinfo {author} {\bibfnamefont {K.}~\bibnamefont {Watanabe}},
  \bibinfo {author} {\bibfnamefont {T.}~\bibnamefont {Taniguchi}}, \bibinfo
  {author} {\bibfnamefont {A.}~\bibnamefont {Zettl}}, \bibinfo {author}
  {\bibfnamefont {S.}~\bibnamefont {Adam}}, \bibinfo {author} {\bibfnamefont
  {A.~H.}\ \bibnamefont {MacDonald}},\ and\ \bibinfo {author} {\bibfnamefont
  {M.~F.}\ \bibnamefont {Crommie}},\ }\bibfield  {title} {\bibinfo {title}
  {Local spectroscopy of moir\'e-induced electronic structure in gate-tunable
  twisted bilayer graphene},\ }\href
  {https://doi.org/10.1103/PhysRevB.92.155409} {\bibfield  {journal} {\bibinfo
  {journal} {Phys. Rev. B}\ }\textbf {\bibinfo {volume} {92}},\ \bibinfo
  {pages} {155409} (\bibinfo {year} {2015})}\BibitemShut {NoStop}%
\bibitem [{\citenamefont {Kim}\ \emph {et~al.}(2017)\citenamefont {Kim},
  \citenamefont {DaSilva}, \citenamefont {Huang}, \citenamefont {Fallahazad},
  \citenamefont {Larentis}, \citenamefont {Taniguchi}, \citenamefont
  {Watanabe}, \citenamefont {LeRoy}, \citenamefont {MacDonald},\ and\
  \citenamefont {Tutuc}}]{Kim3364}%
  \BibitemOpen
  \bibfield  {author} {\bibinfo {author} {\bibfnamefont {K.}~\bibnamefont
  {Kim}}, \bibinfo {author} {\bibfnamefont {A.}~\bibnamefont {DaSilva}},
  \bibinfo {author} {\bibfnamefont {S.}~\bibnamefont {Huang}}, \bibinfo
  {author} {\bibfnamefont {B.}~\bibnamefont {Fallahazad}}, \bibinfo {author}
  {\bibfnamefont {S.}~\bibnamefont {Larentis}}, \bibinfo {author}
  {\bibfnamefont {T.}~\bibnamefont {Taniguchi}}, \bibinfo {author}
  {\bibfnamefont {K.}~\bibnamefont {Watanabe}}, \bibinfo {author}
  {\bibfnamefont {B.~J.}\ \bibnamefont {LeRoy}}, \bibinfo {author}
  {\bibfnamefont {A.~H.}\ \bibnamefont {MacDonald}},\ and\ \bibinfo {author}
  {\bibfnamefont {E.}~\bibnamefont {Tutuc}},\ }\bibfield  {title} {\bibinfo
  {title} {Tunable moir{\'e} bands and strong correlations in small-twist-angle
  bilayer graphene},\ }\href {https://doi.org/10.1073/pnas.1620140114}
  {\bibfield  {journal} {\bibinfo  {journal} {Proc. Natl. Acad. Sci. U.S.A.}\
  }\textbf {\bibinfo {volume} {114}},\ \bibinfo {pages} {3364} (\bibinfo {year}
  {2017})}\BibitemShut {NoStop}%
\bibitem [{\citenamefont {Cao}\ \emph {et~al.}(2018{\natexlab{a}})\citenamefont
  {Cao}, \citenamefont {Fatemi}, \citenamefont {Demir}, \citenamefont {Fang},
  \citenamefont {Tomarken}, \citenamefont {Luo}, \citenamefont
  {Sanchez-Yamagishi}, \citenamefont {Watanabe}, \citenamefont {Taniguchi},
  \citenamefont {Kaxiras}, \citenamefont {Ashoori},\ and\ \citenamefont
  {Jarillo-Herrero}}]{Cao:2018tp}%
  \BibitemOpen
  \bibfield  {author} {\bibinfo {author} {\bibfnamefont {Y.}~\bibnamefont
  {Cao}}, \bibinfo {author} {\bibfnamefont {V.}~\bibnamefont {Fatemi}},
  \bibinfo {author} {\bibfnamefont {A.}~\bibnamefont {Demir}}, \bibinfo
  {author} {\bibfnamefont {S.}~\bibnamefont {Fang}}, \bibinfo {author}
  {\bibfnamefont {S.~L.}\ \bibnamefont {Tomarken}}, \bibinfo {author}
  {\bibfnamefont {J.~Y.}\ \bibnamefont {Luo}}, \bibinfo {author} {\bibfnamefont
  {J.~D.}\ \bibnamefont {Sanchez-Yamagishi}}, \bibinfo {author} {\bibfnamefont
  {K.}~\bibnamefont {Watanabe}}, \bibinfo {author} {\bibfnamefont
  {T.}~\bibnamefont {Taniguchi}}, \bibinfo {author} {\bibfnamefont
  {E.}~\bibnamefont {Kaxiras}}, \bibinfo {author} {\bibfnamefont {R.~C.}\
  \bibnamefont {Ashoori}},\ and\ \bibinfo {author} {\bibfnamefont
  {P.}~\bibnamefont {Jarillo-Herrero}},\ }\bibfield  {title} {\bibinfo {title}
  {Correlated insulator behaviour at half-filling in magic-angle graphene
  superlattices},\ }\href {https://doi.org/10.1038/nature26154} {\bibfield
  {journal} {\bibinfo  {journal} {Nature}\ }\textbf {\bibinfo {volume} {556}},\
  \bibinfo {pages} {80} (\bibinfo {year} {2018}{\natexlab{a}})}\BibitemShut
  {NoStop}%
\bibitem [{\citenamefont {Cao}\ \emph {et~al.}(2018{\natexlab{b}})\citenamefont
  {Cao}, \citenamefont {Fatemi}, \citenamefont {Fang}, \citenamefont
  {Watanabe}, \citenamefont {Taniguchi}, \citenamefont {Kaxiras},\ and\
  \citenamefont {Jarillo-Herrero}}]{Cao:2018wy}%
  \BibitemOpen
  \bibfield  {author} {\bibinfo {author} {\bibfnamefont {Y.}~\bibnamefont
  {Cao}}, \bibinfo {author} {\bibfnamefont {V.}~\bibnamefont {Fatemi}},
  \bibinfo {author} {\bibfnamefont {S.}~\bibnamefont {Fang}}, \bibinfo {author}
  {\bibfnamefont {K.}~\bibnamefont {Watanabe}}, \bibinfo {author}
  {\bibfnamefont {T.}~\bibnamefont {Taniguchi}}, \bibinfo {author}
  {\bibfnamefont {E.}~\bibnamefont {Kaxiras}},\ and\ \bibinfo {author}
  {\bibfnamefont {P.}~\bibnamefont {Jarillo-Herrero}},\ }\bibfield  {title}
  {\bibinfo {title} {Unconventional superconductivity in magic-angle graphene
  superlattices},\ }\href {https://doi.org/10.1038/nature26160} {\bibfield
  {journal} {\bibinfo  {journal} {Nature}\ }\textbf {\bibinfo {volume} {556}},\
  \bibinfo {pages} {43} (\bibinfo {year} {2018}{\natexlab{b}})}\BibitemShut
  {NoStop}%
\bibitem [{\citenamefont {Koshino}\ \emph {et~al.}(2018)\citenamefont
  {Koshino}, \citenamefont {Yuan}, \citenamefont {Koretsune}, \citenamefont
  {Ochi}, \citenamefont {Kuroki},\ and\ \citenamefont
  {Fu}}]{PhysRevX.8.031087}%
  \BibitemOpen
  \bibfield  {author} {\bibinfo {author} {\bibfnamefont {M.}~\bibnamefont
  {Koshino}}, \bibinfo {author} {\bibfnamefont {N.~F.~Q.}\ \bibnamefont
  {Yuan}}, \bibinfo {author} {\bibfnamefont {T.}~\bibnamefont {Koretsune}},
  \bibinfo {author} {\bibfnamefont {M.}~\bibnamefont {Ochi}}, \bibinfo {author}
  {\bibfnamefont {K.}~\bibnamefont {Kuroki}},\ and\ \bibinfo {author}
  {\bibfnamefont {L.}~\bibnamefont {Fu}},\ }\bibfield  {title} {\bibinfo
  {title} {Maximally localized {Wannier} orbitals and the extended {Hubbard}
  model for twisted bilayer graphene},\ }\href
  {https://doi.org/10.1103/PhysRevX.8.031087} {\bibfield  {journal} {\bibinfo
  {journal} {Phys. Rev. X}\ }\textbf {\bibinfo {volume} {8}},\ \bibinfo {pages}
  {031087} (\bibinfo {year} {2018})}\BibitemShut {NoStop}%
\bibitem [{\citenamefont {Sharpe}\ \emph {et~al.}(2019)\citenamefont {Sharpe},
  \citenamefont {Fox}, \citenamefont {Barnard}, \citenamefont {Finney},
  \citenamefont {Watanabe}, \citenamefont {Taniguchi}, \citenamefont
  {Kastner},\ and\ \citenamefont
  {Goldhaber-Gordon}}]{doi:10.1126/science.aaw3780}%
  \BibitemOpen
  \bibfield  {author} {\bibinfo {author} {\bibfnamefont {A.~L.}\ \bibnamefont
  {Sharpe}}, \bibinfo {author} {\bibfnamefont {E.~J.}\ \bibnamefont {Fox}},
  \bibinfo {author} {\bibfnamefont {A.~W.}\ \bibnamefont {Barnard}}, \bibinfo
  {author} {\bibfnamefont {J.}~\bibnamefont {Finney}}, \bibinfo {author}
  {\bibfnamefont {K.}~\bibnamefont {Watanabe}}, \bibinfo {author}
  {\bibfnamefont {T.}~\bibnamefont {Taniguchi}}, \bibinfo {author}
  {\bibfnamefont {M.~A.}\ \bibnamefont {Kastner}},\ and\ \bibinfo {author}
  {\bibfnamefont {D.}~\bibnamefont {Goldhaber-Gordon}},\ }\bibfield  {title}
  {\bibinfo {title} {Emergent ferromagnetism near three-quarters filling in
  twisted bilayer graphene},\ }\href {https://doi.org/10.1126/science.aaw3780}
  {\bibfield  {journal} {\bibinfo  {journal} {Science}\ }\textbf {\bibinfo
  {volume} {365}},\ \bibinfo {pages} {605} (\bibinfo {year}
  {2019})}\BibitemShut {NoStop}%
\bibitem [{\citenamefont {Lu}\ \emph {et~al.}(2019)\citenamefont {Lu},
  \citenamefont {Stepanov}, \citenamefont {Yang}, \citenamefont {Xie},
  \citenamefont {Aamir}, \citenamefont {Das}, \citenamefont {Urgell},
  \citenamefont {Watanabe}, \citenamefont {Taniguchi}, \citenamefont {Zhang},
  \citenamefont {Bachtold}, \citenamefont {MacDonald},\ and\ \citenamefont
  {Efetov}}]{Lu:2019ww}%
  \BibitemOpen
  \bibfield  {author} {\bibinfo {author} {\bibfnamefont {X.}~\bibnamefont
  {Lu}}, \bibinfo {author} {\bibfnamefont {P.}~\bibnamefont {Stepanov}},
  \bibinfo {author} {\bibfnamefont {W.}~\bibnamefont {Yang}}, \bibinfo {author}
  {\bibfnamefont {M.}~\bibnamefont {Xie}}, \bibinfo {author} {\bibfnamefont
  {M.~A.}\ \bibnamefont {Aamir}}, \bibinfo {author} {\bibfnamefont
  {I.}~\bibnamefont {Das}}, \bibinfo {author} {\bibfnamefont {C.}~\bibnamefont
  {Urgell}}, \bibinfo {author} {\bibfnamefont {K.}~\bibnamefont {Watanabe}},
  \bibinfo {author} {\bibfnamefont {T.}~\bibnamefont {Taniguchi}}, \bibinfo
  {author} {\bibfnamefont {G.}~\bibnamefont {Zhang}}, \bibinfo {author}
  {\bibfnamefont {A.}~\bibnamefont {Bachtold}}, \bibinfo {author}
  {\bibfnamefont {A.~H.}\ \bibnamefont {MacDonald}},\ and\ \bibinfo {author}
  {\bibfnamefont {D.~K.}\ \bibnamefont {Efetov}},\ }\bibfield  {title}
  {\bibinfo {title} {Superconductors, orbital magnets and correlated states in
  magic-angle bilayer graphene},\ }\href
  {https://doi.org/10.1038/s41586-019-1695-0} {\bibfield  {journal} {\bibinfo
  {journal} {Nature}\ }\textbf {\bibinfo {volume} {574}},\ \bibinfo {pages}
  {653} (\bibinfo {year} {2019})}\BibitemShut {NoStop}%
\bibitem [{\citenamefont {Yankowitz}\ \emph {et~al.}(2019)\citenamefont
  {Yankowitz}, \citenamefont {Chen}, \citenamefont {Polshyn}, \citenamefont
  {Zhang}, \citenamefont {Watanabe}, \citenamefont {Taniguchi}, \citenamefont
  {Graf}, \citenamefont {Young},\ and\ \citenamefont
  {Dean}}]{doi:10.1126/science.aav1910}%
  \BibitemOpen
  \bibfield  {author} {\bibinfo {author} {\bibfnamefont {M.}~\bibnamefont
  {Yankowitz}}, \bibinfo {author} {\bibfnamefont {S.}~\bibnamefont {Chen}},
  \bibinfo {author} {\bibfnamefont {H.}~\bibnamefont {Polshyn}}, \bibinfo
  {author} {\bibfnamefont {Y.}~\bibnamefont {Zhang}}, \bibinfo {author}
  {\bibfnamefont {K.}~\bibnamefont {Watanabe}}, \bibinfo {author}
  {\bibfnamefont {T.}~\bibnamefont {Taniguchi}}, \bibinfo {author}
  {\bibfnamefont {D.}~\bibnamefont {Graf}}, \bibinfo {author} {\bibfnamefont
  {A.~F.}\ \bibnamefont {Young}},\ and\ \bibinfo {author} {\bibfnamefont
  {C.~R.}\ \bibnamefont {Dean}},\ }\bibfield  {title} {\bibinfo {title} {Tuning
  superconductivity in twisted bilayer graphene},\ }\href
  {https://doi.org/10.1126/science.aav1910} {\bibfield  {journal} {\bibinfo
  {journal} {Science}\ }\textbf {\bibinfo {volume} {363}},\ \bibinfo {pages}
  {1059} (\bibinfo {year} {2019})}\BibitemShut {NoStop}%
\bibitem [{\citenamefont {Wong}\ \emph {et~al.}(2020)\citenamefont {Wong},
  \citenamefont {Nuckolls}, \citenamefont {Oh}, \citenamefont {Lian},
  \citenamefont {Xie}, \citenamefont {Jeon}, \citenamefont {Watanabe},
  \citenamefont {Taniguchi}, \citenamefont {Bernevig},\ and\ \citenamefont
  {Yazdani}}]{Wong:2020vy}%
  \BibitemOpen
  \bibfield  {author} {\bibinfo {author} {\bibfnamefont {D.}~\bibnamefont
  {Wong}}, \bibinfo {author} {\bibfnamefont {K.~P.}\ \bibnamefont {Nuckolls}},
  \bibinfo {author} {\bibfnamefont {M.}~\bibnamefont {Oh}}, \bibinfo {author}
  {\bibfnamefont {B.}~\bibnamefont {Lian}}, \bibinfo {author} {\bibfnamefont
  {Y.}~\bibnamefont {Xie}}, \bibinfo {author} {\bibfnamefont {S.}~\bibnamefont
  {Jeon}}, \bibinfo {author} {\bibfnamefont {K.}~\bibnamefont {Watanabe}},
  \bibinfo {author} {\bibfnamefont {T.}~\bibnamefont {Taniguchi}}, \bibinfo
  {author} {\bibfnamefont {B.~A.}\ \bibnamefont {Bernevig}},\ and\ \bibinfo
  {author} {\bibfnamefont {A.}~\bibnamefont {Yazdani}},\ }\bibfield  {title}
  {\bibinfo {title} {Cascade of electronic transitions in magic-angle twisted
  bilayer graphene},\ }\href {https://doi.org/10.1038/s41586-020-2339-0}
  {\bibfield  {journal} {\bibinfo  {journal} {Nature}\ }\textbf {\bibinfo
  {volume} {582}},\ \bibinfo {pages} {198} (\bibinfo {year}
  {2020})}\BibitemShut {NoStop}%
\bibitem [{\citenamefont {Zondiner}\ \emph {et~al.}(2020)\citenamefont
  {Zondiner}, \citenamefont {Rozen}, \citenamefont {Rodan-Legrain},
  \citenamefont {Cao}, \citenamefont {Queiroz}, \citenamefont {Taniguchi},
  \citenamefont {Watanabe}, \citenamefont {Oreg}, \citenamefont {von Oppen},
  \citenamefont {Stern}, \citenamefont {Berg}, \citenamefont
  {Jarillo-Herrero},\ and\ \citenamefont {Ilani}}]{Zondiner:2020vo}%
  \BibitemOpen
  \bibfield  {author} {\bibinfo {author} {\bibfnamefont {U.}~\bibnamefont
  {Zondiner}}, \bibinfo {author} {\bibfnamefont {A.}~\bibnamefont {Rozen}},
  \bibinfo {author} {\bibfnamefont {D.}~\bibnamefont {Rodan-Legrain}}, \bibinfo
  {author} {\bibfnamefont {Y.}~\bibnamefont {Cao}}, \bibinfo {author}
  {\bibfnamefont {R.}~\bibnamefont {Queiroz}}, \bibinfo {author} {\bibfnamefont
  {T.}~\bibnamefont {Taniguchi}}, \bibinfo {author} {\bibfnamefont
  {K.}~\bibnamefont {Watanabe}}, \bibinfo {author} {\bibfnamefont
  {Y.}~\bibnamefont {Oreg}}, \bibinfo {author} {\bibfnamefont {F.}~\bibnamefont
  {von Oppen}}, \bibinfo {author} {\bibfnamefont {A.}~\bibnamefont {Stern}},
  \bibinfo {author} {\bibfnamefont {E.}~\bibnamefont {Berg}}, \bibinfo {author}
  {\bibfnamefont {P.}~\bibnamefont {Jarillo-Herrero}},\ and\ \bibinfo {author}
  {\bibfnamefont {S.}~\bibnamefont {Ilani}},\ }\bibfield  {title} {\bibinfo
  {title} {Cascade of phase transitions and {Dirac} revivals in magic-angle
  graphene},\ }\href {https://doi.org/10.1038/s41586-020-2373-y} {\bibfield
  {journal} {\bibinfo  {journal} {Nature}\ }\textbf {\bibinfo {volume} {582}},\
  \bibinfo {pages} {203} (\bibinfo {year} {2020})}\BibitemShut {NoStop}%
\bibitem [{\citenamefont {Serlin}\ \emph {et~al.}(2020)\citenamefont {Serlin},
  \citenamefont {Tschirhart}, \citenamefont {Polshyn}, \citenamefont {Zhang},
  \citenamefont {Zhu}, \citenamefont {Watanabe}, \citenamefont {Taniguchi},
  \citenamefont {Balents},\ and\ \citenamefont
  {Young}}]{doi:10.1126/science.aay5533}%
  \BibitemOpen
  \bibfield  {author} {\bibinfo {author} {\bibfnamefont {M.}~\bibnamefont
  {Serlin}}, \bibinfo {author} {\bibfnamefont {C.~L.}\ \bibnamefont
  {Tschirhart}}, \bibinfo {author} {\bibfnamefont {H.}~\bibnamefont {Polshyn}},
  \bibinfo {author} {\bibfnamefont {Y.}~\bibnamefont {Zhang}}, \bibinfo
  {author} {\bibfnamefont {J.}~\bibnamefont {Zhu}}, \bibinfo {author}
  {\bibfnamefont {K.}~\bibnamefont {Watanabe}}, \bibinfo {author}
  {\bibfnamefont {T.}~\bibnamefont {Taniguchi}}, \bibinfo {author}
  {\bibfnamefont {L.}~\bibnamefont {Balents}},\ and\ \bibinfo {author}
  {\bibfnamefont {A.~F.}\ \bibnamefont {Young}},\ }\bibfield  {title} {\bibinfo
  {title} {Intrinsic quantized anomalous hall effect in a moir\'e
  heterostructure},\ }\href {https://doi.org/10.1126/science.aay5533}
  {\bibfield  {journal} {\bibinfo  {journal} {Science}\ }\textbf {\bibinfo
  {volume} {367}},\ \bibinfo {pages} {900} (\bibinfo {year}
  {2020})}\BibitemShut {NoStop}%
\bibitem [{\citenamefont {Nuckolls}\ \emph {et~al.}(2020)\citenamefont
  {Nuckolls}, \citenamefont {Oh}, \citenamefont {Wong}, \citenamefont {Lian},
  \citenamefont {Watanabe}, \citenamefont {Taniguchi}, \citenamefont
  {Bernevig},\ and\ \citenamefont {Yazdani}}]{Nuckolls:2020wp}%
  \BibitemOpen
  \bibfield  {author} {\bibinfo {author} {\bibfnamefont {K.~P.}\ \bibnamefont
  {Nuckolls}}, \bibinfo {author} {\bibfnamefont {M.}~\bibnamefont {Oh}},
  \bibinfo {author} {\bibfnamefont {D.}~\bibnamefont {Wong}}, \bibinfo {author}
  {\bibfnamefont {B.}~\bibnamefont {Lian}}, \bibinfo {author} {\bibfnamefont
  {K.}~\bibnamefont {Watanabe}}, \bibinfo {author} {\bibfnamefont
  {T.}~\bibnamefont {Taniguchi}}, \bibinfo {author} {\bibfnamefont {B.~A.}\
  \bibnamefont {Bernevig}},\ and\ \bibinfo {author} {\bibfnamefont
  {A.}~\bibnamefont {Yazdani}},\ }\bibfield  {title} {\bibinfo {title}
  {Strongly correlated {Chern} insulators in magic-angle twisted bilayer
  graphene},\ }\href {https://doi.org/10.1038/s41586-020-3028-8} {\bibfield
  {journal} {\bibinfo  {journal} {Nature}\ }\textbf {\bibinfo {volume} {588}},\
  \bibinfo {pages} {610} (\bibinfo {year} {2020})}\BibitemShut {NoStop}%
\bibitem [{\citenamefont {Choi}\ \emph {et~al.}(2021)\citenamefont {Choi},
  \citenamefont {Kim}, \citenamefont {Peng}, \citenamefont {Thomson},
  \citenamefont {Lewandowski}, \citenamefont {Polski}, \citenamefont {Zhang},
  \citenamefont {Arora}, \citenamefont {Watanabe}, \citenamefont {Taniguchi},
  \citenamefont {Alicea},\ and\ \citenamefont {Nadj-Perge}}]{Choi:2021vi}%
  \BibitemOpen
  \bibfield  {author} {\bibinfo {author} {\bibfnamefont {Y.}~\bibnamefont
  {Choi}}, \bibinfo {author} {\bibfnamefont {H.}~\bibnamefont {Kim}}, \bibinfo
  {author} {\bibfnamefont {Y.}~\bibnamefont {Peng}}, \bibinfo {author}
  {\bibfnamefont {A.}~\bibnamefont {Thomson}}, \bibinfo {author} {\bibfnamefont
  {C.}~\bibnamefont {Lewandowski}}, \bibinfo {author} {\bibfnamefont
  {R.}~\bibnamefont {Polski}}, \bibinfo {author} {\bibfnamefont
  {Y.}~\bibnamefont {Zhang}}, \bibinfo {author} {\bibfnamefont {H.~S.}\
  \bibnamefont {Arora}}, \bibinfo {author} {\bibfnamefont {K.}~\bibnamefont
  {Watanabe}}, \bibinfo {author} {\bibfnamefont {T.}~\bibnamefont {Taniguchi}},
  \bibinfo {author} {\bibfnamefont {J.}~\bibnamefont {Alicea}},\ and\ \bibinfo
  {author} {\bibfnamefont {S.}~\bibnamefont {Nadj-Perge}},\ }\bibfield  {title}
  {\bibinfo {title} {Correlation-driven topological phases in magic-angle
  twisted bilayer graphene},\ }\href
  {https://doi.org/10.1038/s41586-020-03159-7} {\bibfield  {journal} {\bibinfo
  {journal} {Nature}\ }\textbf {\bibinfo {volume} {589}},\ \bibinfo {pages}
  {536} (\bibinfo {year} {2021})}\BibitemShut {NoStop}%
\bibitem [{\citenamefont {Kariyado}\ and\ \citenamefont
  {Vishwanath}(2019)}]{PhysRevResearch.1.033076}%
  \BibitemOpen
  \bibfield  {author} {\bibinfo {author} {\bibfnamefont {T.}~\bibnamefont
  {Kariyado}}\ and\ \bibinfo {author} {\bibfnamefont {A.}~\bibnamefont
  {Vishwanath}},\ }\bibfield  {title} {\bibinfo {title} {Flat band in twisted
  bilayer {Bravais} lattices},\ }\href
  {https://doi.org/10.1103/PhysRevResearch.1.033076} {\bibfield  {journal}
  {\bibinfo  {journal} {Phys. Rev. Research}\ }\textbf {\bibinfo {volume}
  {1}},\ \bibinfo {pages} {033076} (\bibinfo {year} {2019})}\BibitemShut
  {NoStop}%
\bibitem [{\citenamefont {Rycerz}\ \emph {et~al.}(2007)\citenamefont {Rycerz},
  \citenamefont {Tworzydlo},\ and\ \citenamefont {Beenakker}}]{Rycerz:2007vn}%
  \BibitemOpen
  \bibfield  {author} {\bibinfo {author} {\bibfnamefont {A.}~\bibnamefont
  {Rycerz}}, \bibinfo {author} {\bibfnamefont {J.}~\bibnamefont {Tworzydlo}},\
  and\ \bibinfo {author} {\bibfnamefont {C.~W.~J.}\ \bibnamefont {Beenakker}},\
  }\bibfield  {title} {\bibinfo {title} {Valley filter and valley valve in
  graphene},\ }\href {http://dx.doi.org/10.1038/nphys547} {\bibfield  {journal}
  {\bibinfo  {journal} {Nat. Phys.}\ }\textbf {\bibinfo {volume} {3}},\
  \bibinfo {pages} {172} (\bibinfo {year} {2007})}\BibitemShut {NoStop}%
\bibitem [{\citenamefont {Tanaka}\ \emph {et~al.}(2005)\citenamefont {Tanaka},
  \citenamefont {Kawamata}, \citenamefont {Simizu}, \citenamefont {Fujita},
  \citenamefont {Yanagisawa}, \citenamefont {Otani},\ and\ \citenamefont
  {Oshima}}]{TANAKA200522}%
  \BibitemOpen
  \bibfield  {author} {\bibinfo {author} {\bibfnamefont {H.}~\bibnamefont
  {Tanaka}}, \bibinfo {author} {\bibfnamefont {Y.}~\bibnamefont {Kawamata}},
  \bibinfo {author} {\bibfnamefont {H.}~\bibnamefont {Simizu}}, \bibinfo
  {author} {\bibfnamefont {T.}~\bibnamefont {Fujita}}, \bibinfo {author}
  {\bibfnamefont {H.}~\bibnamefont {Yanagisawa}}, \bibinfo {author}
  {\bibfnamefont {S.}~\bibnamefont {Otani}},\ and\ \bibinfo {author}
  {\bibfnamefont {C.}~\bibnamefont {Oshima}},\ }\bibfield  {title} {\bibinfo
  {title} {Novel macroscopic {BC$_3$} honeycomb sheet},\ }\href
  {https://doi.org/https://doi.org/10.1016/j.ssc.2005.06.025} {\bibfield
  {journal} {\bibinfo  {journal} {Solid State Commun.}\ }\textbf {\bibinfo
  {volume} {136}},\ \bibinfo {pages} {22} (\bibinfo {year} {2005})}\BibitemShut
  {NoStop}%
\bibitem [{\citenamefont {Yanagisawa}\ \emph {et~al.}(2006)\citenamefont
  {Yanagisawa}, \citenamefont {Ishida}, \citenamefont {Tanaka}, \citenamefont
  {Ueno}, \citenamefont {Otani},\ and\ \citenamefont
  {Oshima}}]{YANAGISAWA20064072}%
  \BibitemOpen
  \bibfield  {author} {\bibinfo {author} {\bibfnamefont {H.}~\bibnamefont
  {Yanagisawa}}, \bibinfo {author} {\bibfnamefont {Y.}~\bibnamefont {Ishida}},
  \bibinfo {author} {\bibfnamefont {T.}~\bibnamefont {Tanaka}}, \bibinfo
  {author} {\bibfnamefont {A.}~\bibnamefont {Ueno}}, \bibinfo {author}
  {\bibfnamefont {S.}~\bibnamefont {Otani}},\ and\ \bibinfo {author}
  {\bibfnamefont {C.}~\bibnamefont {Oshima}},\ }\bibfield  {title} {\bibinfo
  {title} {Metastable {BC$_3$} honeycomb epitaxial sheets on the
  {NbB$_2$(0001)} surface},\ }\href
  {https://doi.org/https://doi.org/10.1016/j.susc.2006.01.124} {\bibfield
  {journal} {\bibinfo  {journal} {Surf. Sci.}\ }\textbf {\bibinfo {volume}
  {600}},\ \bibinfo {pages} {4072} (\bibinfo {year} {2006})}\BibitemShut
  {NoStop}%
\bibitem [{\citenamefont {Yanagisawa}\ \emph {et~al.}(2004)\citenamefont
  {Yanagisawa}, \citenamefont {Tanaka}, \citenamefont {Ishida}, \citenamefont
  {Matsue}, \citenamefont {Rokuta}, \citenamefont {Otani},\ and\ \citenamefont
  {Oshima}}]{PhysRevLett.93.177003}%
  \BibitemOpen
  \bibfield  {author} {\bibinfo {author} {\bibfnamefont {H.}~\bibnamefont
  {Yanagisawa}}, \bibinfo {author} {\bibfnamefont {T.}~\bibnamefont {Tanaka}},
  \bibinfo {author} {\bibfnamefont {Y.}~\bibnamefont {Ishida}}, \bibinfo
  {author} {\bibfnamefont {M.}~\bibnamefont {Matsue}}, \bibinfo {author}
  {\bibfnamefont {E.}~\bibnamefont {Rokuta}}, \bibinfo {author} {\bibfnamefont
  {S.}~\bibnamefont {Otani}},\ and\ \bibinfo {author} {\bibfnamefont
  {C.}~\bibnamefont {Oshima}},\ }\bibfield  {title} {\bibinfo {title} {Phonon
  dispersion curves of a {${\mathrm{B}\mathrm{C}}_{3}$} honeycomb epitaxial
  sheet},\ }\href {https://doi.org/10.1103/PhysRevLett.93.177003} {\bibfield
  {journal} {\bibinfo  {journal} {Phys. Rev. Lett.}\ }\textbf {\bibinfo
  {volume} {93}},\ \bibinfo {pages} {177003} (\bibinfo {year}
  {2004})}\BibitemShut {NoStop}%
\bibitem [{\citenamefont {Ueno}\ \emph {et~al.}(2006)\citenamefont {Ueno},
  \citenamefont {Fujita}, \citenamefont {Matsue}, \citenamefont {Yanagisawa},
  \citenamefont {Oshima}, \citenamefont {Patthey}, \citenamefont {Ploigt},
  \citenamefont {Schneider},\ and\ \citenamefont {Otani}}]{UENO20063518}%
  \BibitemOpen
  \bibfield  {author} {\bibinfo {author} {\bibfnamefont {A.}~\bibnamefont
  {Ueno}}, \bibinfo {author} {\bibfnamefont {T.}~\bibnamefont {Fujita}},
  \bibinfo {author} {\bibfnamefont {M.}~\bibnamefont {Matsue}}, \bibinfo
  {author} {\bibfnamefont {H.}~\bibnamefont {Yanagisawa}}, \bibinfo {author}
  {\bibfnamefont {C.}~\bibnamefont {Oshima}}, \bibinfo {author} {\bibfnamefont
  {F.}~\bibnamefont {Patthey}}, \bibinfo {author} {\bibfnamefont {H.-C.}\
  \bibnamefont {Ploigt}}, \bibinfo {author} {\bibfnamefont {W.-D.}\
  \bibnamefont {Schneider}},\ and\ \bibinfo {author} {\bibfnamefont
  {S.}~\bibnamefont {Otani}},\ }\bibfield  {title} {\bibinfo {title} {Scanning
  tunneling microscopy study on a {BC$_3$} covered {NbB$_2$(0001)} surface},\
  }\href {https://doi.org/https://doi.org/10.1016/j.susc.2006.07.007}
  {\bibfield  {journal} {\bibinfo  {journal} {Surf. Sci.}\ }\textbf {\bibinfo
  {volume} {600}},\ \bibinfo {pages} {3518} (\bibinfo {year}
  {2006})}\BibitemShut {NoStop}%
\bibitem [{\citenamefont {Wentzcovitch}\ \emph {et~al.}(1988)\citenamefont
  {Wentzcovitch}, \citenamefont {Cohen}, \citenamefont {Louie},\ and\
  \citenamefont {Tománek}}]{WENTZCOVITCH1988515}%
  \BibitemOpen
  \bibfield  {author} {\bibinfo {author} {\bibfnamefont {R.~M.}\ \bibnamefont
  {Wentzcovitch}}, \bibinfo {author} {\bibfnamefont {M.~L.}\ \bibnamefont
  {Cohen}}, \bibinfo {author} {\bibfnamefont {S.~G.}\ \bibnamefont {Louie}},\
  and\ \bibinfo {author} {\bibfnamefont {D.}~\bibnamefont {Tománek}},\
  }\bibfield  {title} {\bibinfo {title} {$\sigma$-states contribution to the
  conductivity of {BC$_3$}},\ }\href
  {https://doi.org/https://doi.org/10.1016/0038-1098(84)90173-X} {\bibfield
  {journal} {\bibinfo  {journal} {Solid State Commun.}\ }\textbf {\bibinfo
  {volume} {67}},\ \bibinfo {pages} {515} (\bibinfo {year} {1988})}\BibitemShut
  {NoStop}%
\bibitem [{\citenamefont {Miyamoto}\ \emph {et~al.}(1994)\citenamefont
  {Miyamoto}, \citenamefont {Rubio}, \citenamefont {Louie},\ and\ \citenamefont
  {Cohen}}]{PhysRevB.50.18360}%
  \BibitemOpen
  \bibfield  {author} {\bibinfo {author} {\bibfnamefont {Y.}~\bibnamefont
  {Miyamoto}}, \bibinfo {author} {\bibfnamefont {A.}~\bibnamefont {Rubio}},
  \bibinfo {author} {\bibfnamefont {S.~G.}\ \bibnamefont {Louie}},\ and\
  \bibinfo {author} {\bibfnamefont {M.~L.}\ \bibnamefont {Cohen}},\ }\bibfield
  {title} {\bibinfo {title} {Electronic properties of tubule forms of hexagonal
  {${\mathrm{BC}}_{3}$}},\ }\href {https://doi.org/10.1103/PhysRevB.50.18360}
  {\bibfield  {journal} {\bibinfo  {journal} {Phys. Rev. B}\ }\textbf {\bibinfo
  {volume} {50}},\ \bibinfo {pages} {18360} (\bibinfo {year}
  {1994})}\BibitemShut {NoStop}%
\bibitem [{\citenamefont {Behzad}(2017)}]{BEHZAD201737}%
  \BibitemOpen
  \bibfield  {author} {\bibinfo {author} {\bibfnamefont {S.}~\bibnamefont
  {Behzad}},\ }\bibfield  {title} {\bibinfo {title} {Mechanical control of the
  electro-optical properties of monolayer and bilayer {BC$_3$} by applying the
  in-plane biaxial strain},\ }\href
  {https://doi.org/https://doi.org/10.1016/j.susc.2017.07.005} {\bibfield
  {journal} {\bibinfo  {journal} {Surf. Sci.}\ }\textbf {\bibinfo {volume}
  {665}},\ \bibinfo {pages} {37} (\bibinfo {year} {2017})}\BibitemShut
  {NoStop}%
\bibitem [{\citenamefont {Zhang}\ \emph {et~al.}(2018)\citenamefont {Zhang},
  \citenamefont {Liao}, \citenamefont {Yang},\ and\ \citenamefont
  {Zhou}}]{doi:10.1021/acsomega.8b01998}%
  \BibitemOpen
  \bibfield  {author} {\bibinfo {author} {\bibfnamefont {H.}~\bibnamefont
  {Zhang}}, \bibinfo {author} {\bibfnamefont {Y.}~\bibnamefont {Liao}},
  \bibinfo {author} {\bibfnamefont {G.}~\bibnamefont {Yang}},\ and\ \bibinfo
  {author} {\bibfnamefont {X.}~\bibnamefont {Zhou}},\ }\bibfield  {title}
  {\bibinfo {title} {Theoretical studies on the electronic and optical
  properties of honeycomb {BC$_3$} monolayer: A promising candidate for
  metal-free photocatalysts},\ }\href
  {https://doi.org/10.1021/acsomega.8b01998} {\bibfield  {journal} {\bibinfo
  {journal} {ACS Omega}\ }\textbf {\bibinfo {volume} {3}},\ \bibinfo {pages}
  {10517} (\bibinfo {year} {2018})}\BibitemShut {NoStop}%
\bibitem [{\citenamefont {Wang}\ \emph {et~al.}(2019)\citenamefont {Wang},
  \citenamefont {Li}, \citenamefont {Pan}, \citenamefont {Gao},\ and\
  \citenamefont {Sun}}]{doi:10.1063/1.5122678}%
  \BibitemOpen
  \bibfield  {author} {\bibinfo {author} {\bibfnamefont {H.}~\bibnamefont
  {Wang}}, \bibinfo {author} {\bibfnamefont {Q.}~\bibnamefont {Li}}, \bibinfo
  {author} {\bibfnamefont {H.}~\bibnamefont {Pan}}, \bibinfo {author}
  {\bibfnamefont {Y.}~\bibnamefont {Gao}},\ and\ \bibinfo {author}
  {\bibfnamefont {M.}~\bibnamefont {Sun}},\ }\bibfield  {title} {\bibinfo
  {title} {Comparative investigation of the mechanical, electrical and thermal
  transport properties in graphene-like {C$_3$B} and {C$_3$N}},\ }\href
  {https://doi.org/10.1063/1.5122678} {\bibfield  {journal} {\bibinfo
  {journal} {J. Appl. Phys.}\ }\textbf {\bibinfo {volume} {126}},\ \bibinfo
  {pages} {234302} (\bibinfo {year} {2019})}\BibitemShut {NoStop}%
\bibitem [{\citenamefont {Wu}\ \emph {et~al.}(2020)\citenamefont {Wu},
  \citenamefont {Xia}, \citenamefont {Zhang}, \citenamefont {Zhu},
  \citenamefont {Zhang},\ and\ \citenamefont
  {Zhang}}]{PhysRevApplied.14.014073}%
  \BibitemOpen
  \bibfield  {author} {\bibinfo {author} {\bibfnamefont {Y.}~\bibnamefont
  {Wu}}, \bibinfo {author} {\bibfnamefont {W.}~\bibnamefont {Xia}}, \bibinfo
  {author} {\bibfnamefont {Y.}~\bibnamefont {Zhang}}, \bibinfo {author}
  {\bibfnamefont {W.}~\bibnamefont {Zhu}}, \bibinfo {author} {\bibfnamefont
  {W.}~\bibnamefont {Zhang}},\ and\ \bibinfo {author} {\bibfnamefont
  {P.}~\bibnamefont {Zhang}},\ }\bibfield  {title} {\bibinfo {title}
  {Remarkable band-gap renormalization via dimensionality of the layered
  material {${\mathrm{C}}_{3}\mathrm{B}$}},\ }\href
  {https://doi.org/10.1103/PhysRevApplied.14.014073} {\bibfield  {journal}
  {\bibinfo  {journal} {Phys. Rev. Applied}\ }\textbf {\bibinfo {volume}
  {14}},\ \bibinfo {pages} {014073} (\bibinfo {year} {2020})}\BibitemShut
  {NoStop}%
\bibitem [{\citenamefont {Wang}\ \emph {et~al.}(2020)\citenamefont {Wang},
  \citenamefont {Luo}, \citenamefont {Li}, \citenamefont {Yang},\ and\
  \citenamefont {Zhou}}]{D0CP04219F}%
  \BibitemOpen
  \bibfield  {author} {\bibinfo {author} {\bibfnamefont {Z.}~\bibnamefont
  {Wang}}, \bibinfo {author} {\bibfnamefont {Z.}~\bibnamefont {Luo}}, \bibinfo
  {author} {\bibfnamefont {J.}~\bibnamefont {Li}}, \bibinfo {author}
  {\bibfnamefont {K.}~\bibnamefont {Yang}},\ and\ \bibinfo {author}
  {\bibfnamefont {G.}~\bibnamefont {Zhou}},\ }\bibfield  {title} {\bibinfo
  {title} {2d van der {Waals} heterostructures of graphitic {BCN} as direct
  {Z}-scheme photocatalysts for overall water splitting: the role of polar
  $\pi$-conjugated moieties},\ }\href {https://doi.org/10.1039/D0CP04219F}
  {\bibfield  {journal} {\bibinfo  {journal} {Phys. Chem. Chem. Phys.}\
  }\textbf {\bibinfo {volume} {22}},\ \bibinfo {pages} {23735} (\bibinfo {year}
  {2020})}\BibitemShut {NoStop}%
\bibitem [{\citenamefont {Fujimoto}\ and\ \citenamefont
  {Kariyado}(2021)}]{PhysRevB.104.125427}%
  \BibitemOpen
  \bibfield  {author} {\bibinfo {author} {\bibfnamefont {M.}~\bibnamefont
  {Fujimoto}}\ and\ \bibinfo {author} {\bibfnamefont {T.}~\bibnamefont
  {Kariyado}},\ }\bibfield  {title} {\bibinfo {title} {Effective continuum
  model of twisted bilayer {GeSe} and origin of the emerging one-dimensional
  mode},\ }\href {https://doi.org/10.1103/PhysRevB.104.125427} {\bibfield
  {journal} {\bibinfo  {journal} {Phys. Rev. B}\ }\textbf {\bibinfo {volume}
  {104}},\ \bibinfo {pages} {125427} (\bibinfo {year} {2021})}\BibitemShut
  {NoStop}%
\bibitem [{\citenamefont {Jung}\ \emph {et~al.}(2014)\citenamefont {Jung},
  \citenamefont {Raoux}, \citenamefont {Qiao},\ and\ \citenamefont
  {MacDonald}}]{PhysRevB.89.205414}%
  \BibitemOpen
  \bibfield  {author} {\bibinfo {author} {\bibfnamefont {J.}~\bibnamefont
  {Jung}}, \bibinfo {author} {\bibfnamefont {A.}~\bibnamefont {Raoux}},
  \bibinfo {author} {\bibfnamefont {Z.}~\bibnamefont {Qiao}},\ and\ \bibinfo
  {author} {\bibfnamefont {A.~H.}\ \bibnamefont {MacDonald}},\ }\bibfield
  {title} {\bibinfo {title} {Ab initio theory of moir\'e superlattice bands in
  layered two-dimensional materials},\ }\href
  {https://doi.org/10.1103/PhysRevB.89.205414} {\bibfield  {journal} {\bibinfo
  {journal} {Phys. Rev. B}\ }\textbf {\bibinfo {volume} {89}},\ \bibinfo
  {pages} {205414} (\bibinfo {year} {2014})}\BibitemShut {NoStop}%
\bibitem [{\citenamefont {Uchida}\ \emph {et~al.}(2014)\citenamefont {Uchida},
  \citenamefont {Furuya}, \citenamefont {Iwata},\ and\ \citenamefont
  {Oshiyama}}]{PhysRevB.90.155451}%
  \BibitemOpen
  \bibfield  {author} {\bibinfo {author} {\bibfnamefont {K.}~\bibnamefont
  {Uchida}}, \bibinfo {author} {\bibfnamefont {S.}~\bibnamefont {Furuya}},
  \bibinfo {author} {\bibfnamefont {J.-I.}\ \bibnamefont {Iwata}},\ and\
  \bibinfo {author} {\bibfnamefont {A.}~\bibnamefont {Oshiyama}},\ }\bibfield
  {title} {\bibinfo {title} {Atomic corrugation and electron localization due
  to moir\'e patterns in twisted bilayer graphenes},\ }\href
  {https://doi.org/10.1103/PhysRevB.90.155451} {\bibfield  {journal} {\bibinfo
  {journal} {Phys. Rev. B}\ }\textbf {\bibinfo {volume} {90}},\ \bibinfo
  {pages} {155451} (\bibinfo {year} {2014})}\BibitemShut {NoStop}%
\bibitem [{\citenamefont {Akashi}\ \emph {et~al.}(2017)\citenamefont {Akashi},
  \citenamefont {Iida}, \citenamefont {Yamamoto},\ and\ \citenamefont
  {Yoshizawa}}]{PhysRevB.95.245401}%
  \BibitemOpen
  \bibfield  {author} {\bibinfo {author} {\bibfnamefont {R.}~\bibnamefont
  {Akashi}}, \bibinfo {author} {\bibfnamefont {Y.}~\bibnamefont {Iida}},
  \bibinfo {author} {\bibfnamefont {K.}~\bibnamefont {Yamamoto}},\ and\
  \bibinfo {author} {\bibfnamefont {K.}~\bibnamefont {Yoshizawa}},\ }\bibfield
  {title} {\bibinfo {title} {Interference of the {Bloch} phase in layered
  materials with stacking shifts},\ }\href
  {https://doi.org/10.1103/PhysRevB.95.245401} {\bibfield  {journal} {\bibinfo
  {journal} {Phys. Rev. B}\ }\textbf {\bibinfo {volume} {95}},\ \bibinfo
  {pages} {245401} (\bibinfo {year} {2017})}\BibitemShut {NoStop}%
\bibitem [{\citenamefont {Moon}\ and\ \citenamefont
  {Koshino}(2013)}]{PhysRevB.87.205404}%
  \BibitemOpen
  \bibfield  {author} {\bibinfo {author} {\bibfnamefont {P.}~\bibnamefont
  {Moon}}\ and\ \bibinfo {author} {\bibfnamefont {M.}~\bibnamefont {Koshino}},\
  }\bibfield  {title} {\bibinfo {title} {Optical absorption in twisted bilayer
  graphene},\ }\href {https://doi.org/10.1103/PhysRevB.87.205404} {\bibfield
  {journal} {\bibinfo  {journal} {Phys. Rev. B}\ }\textbf {\bibinfo {volume}
  {87}},\ \bibinfo {pages} {205404} (\bibinfo {year} {2013})}\BibitemShut
  {NoStop}%
\bibitem [{\citenamefont {Wang}\ \emph {et~al.}(2017)\citenamefont {Wang},
  \citenamefont {Wang}, \citenamefont {Yao}, \citenamefont {Liu},\ and\
  \citenamefont {Yu}}]{PhysRevB.95.115429}%
  \BibitemOpen
  \bibfield  {author} {\bibinfo {author} {\bibfnamefont {Y.}~\bibnamefont
  {Wang}}, \bibinfo {author} {\bibfnamefont {Z.}~\bibnamefont {Wang}}, \bibinfo
  {author} {\bibfnamefont {W.}~\bibnamefont {Yao}}, \bibinfo {author}
  {\bibfnamefont {G.-B.}\ \bibnamefont {Liu}},\ and\ \bibinfo {author}
  {\bibfnamefont {H.}~\bibnamefont {Yu}},\ }\bibfield  {title} {\bibinfo
  {title} {Interlayer coupling in commensurate and incommensurate bilayer
  structures of transition-metal dichalcogenides},\ }\href
  {https://doi.org/10.1103/PhysRevB.95.115429} {\bibfield  {journal} {\bibinfo
  {journal} {Phys. Rev. B}\ }\textbf {\bibinfo {volume} {95}},\ \bibinfo
  {pages} {115429} (\bibinfo {year} {2017})}\BibitemShut {NoStop}%
\bibitem [{\citenamefont {Rost}\ \emph {et~al.}(2019)\citenamefont {Rost},
  \citenamefont {Gupta}, \citenamefont {Fleischmann}, \citenamefont
  {Weckbecker}, \citenamefont {Ray}, \citenamefont {Olivares}, \citenamefont
  {Vogl}, \citenamefont {Sharma}, \citenamefont {Pankratov},\ and\
  \citenamefont {Shallcross}}]{PhysRevB.100.035101}%
  \BibitemOpen
  \bibfield  {author} {\bibinfo {author} {\bibfnamefont {F.}~\bibnamefont
  {Rost}}, \bibinfo {author} {\bibfnamefont {R.}~\bibnamefont {Gupta}},
  \bibinfo {author} {\bibfnamefont {M.}~\bibnamefont {Fleischmann}}, \bibinfo
  {author} {\bibfnamefont {D.}~\bibnamefont {Weckbecker}}, \bibinfo {author}
  {\bibfnamefont {N.}~\bibnamefont {Ray}}, \bibinfo {author} {\bibfnamefont
  {J.}~\bibnamefont {Olivares}}, \bibinfo {author} {\bibfnamefont
  {M.}~\bibnamefont {Vogl}}, \bibinfo {author} {\bibfnamefont {S.}~\bibnamefont
  {Sharma}}, \bibinfo {author} {\bibfnamefont {O.}~\bibnamefont {Pankratov}},\
  and\ \bibinfo {author} {\bibfnamefont {S.}~\bibnamefont {Shallcross}},\
  }\bibfield  {title} {\bibinfo {title} {Nonperturbative theory of effective
  {Hamiltonians} for deformations in two-dimensional materials: Moir\'e systems
  and dislocations},\ }\href {https://doi.org/10.1103/PhysRevB.100.035101}
  {\bibfield  {journal} {\bibinfo  {journal} {Phys. Rev. B}\ }\textbf {\bibinfo
  {volume} {100}},\ \bibinfo {pages} {035101} (\bibinfo {year}
  {2019})}\BibitemShut {NoStop}%
\bibitem [{\citenamefont {Giannozzi}\ \emph {et~al.}(2009)\citenamefont
  {Giannozzi}, \citenamefont {Baroni}, \citenamefont {Bonini}, \citenamefont
  {Calandra}, \citenamefont {Car}, \citenamefont {Cavazzoni}, \citenamefont
  {Ceresoli}, \citenamefont {Chiarotti}, \citenamefont {Cococcioni},
  \citenamefont {Dabo}, \citenamefont {Corso}, \citenamefont {de~Gironcoli},
  \citenamefont {Fabris}, \citenamefont {Fratesi}, \citenamefont {Gebauer},
  \citenamefont {Gerstmann}, \citenamefont {Gougoussis}, \citenamefont
  {Kokalj}, \citenamefont {Lazzeri}, \citenamefont {Martin-Samos},
  \citenamefont {Marzari}, \citenamefont {Mauri}, \citenamefont {Mazzarello},
  \citenamefont {Paolini}, \citenamefont {Pasquarello}, \citenamefont
  {Paulatto}, \citenamefont {Sbraccia}, \citenamefont {Scandolo}, \citenamefont
  {Sclauzero}, \citenamefont {Seitsonen}, \citenamefont {Smogunov},
  \citenamefont {Umari},\ and\ \citenamefont {Wentzcovitch}}]{Giannozzi_2009}%
  \BibitemOpen
  \bibfield  {author} {\bibinfo {author} {\bibfnamefont {P.}~\bibnamefont
  {Giannozzi}}, \bibinfo {author} {\bibfnamefont {S.}~\bibnamefont {Baroni}},
  \bibinfo {author} {\bibfnamefont {N.}~\bibnamefont {Bonini}}, \bibinfo
  {author} {\bibfnamefont {M.}~\bibnamefont {Calandra}}, \bibinfo {author}
  {\bibfnamefont {R.}~\bibnamefont {Car}}, \bibinfo {author} {\bibfnamefont
  {C.}~\bibnamefont {Cavazzoni}}, \bibinfo {author} {\bibfnamefont
  {D.}~\bibnamefont {Ceresoli}}, \bibinfo {author} {\bibfnamefont {G.~L.}\
  \bibnamefont {Chiarotti}}, \bibinfo {author} {\bibfnamefont {M.}~\bibnamefont
  {Cococcioni}}, \bibinfo {author} {\bibfnamefont {I.}~\bibnamefont {Dabo}},
  \bibinfo {author} {\bibfnamefont {A.~D.}\ \bibnamefont {Corso}}, \bibinfo
  {author} {\bibfnamefont {S.}~\bibnamefont {de~Gironcoli}}, \bibinfo {author}
  {\bibfnamefont {S.}~\bibnamefont {Fabris}}, \bibinfo {author} {\bibfnamefont
  {G.}~\bibnamefont {Fratesi}}, \bibinfo {author} {\bibfnamefont
  {R.}~\bibnamefont {Gebauer}}, \bibinfo {author} {\bibfnamefont
  {U.}~\bibnamefont {Gerstmann}}, \bibinfo {author} {\bibfnamefont
  {C.}~\bibnamefont {Gougoussis}}, \bibinfo {author} {\bibfnamefont
  {A.}~\bibnamefont {Kokalj}}, \bibinfo {author} {\bibfnamefont
  {M.}~\bibnamefont {Lazzeri}}, \bibinfo {author} {\bibfnamefont
  {L.}~\bibnamefont {Martin-Samos}}, \bibinfo {author} {\bibfnamefont
  {N.}~\bibnamefont {Marzari}}, \bibinfo {author} {\bibfnamefont
  {F.}~\bibnamefont {Mauri}}, \bibinfo {author} {\bibfnamefont
  {R.}~\bibnamefont {Mazzarello}}, \bibinfo {author} {\bibfnamefont
  {S.}~\bibnamefont {Paolini}}, \bibinfo {author} {\bibfnamefont
  {A.}~\bibnamefont {Pasquarello}}, \bibinfo {author} {\bibfnamefont
  {L.}~\bibnamefont {Paulatto}}, \bibinfo {author} {\bibfnamefont
  {C.}~\bibnamefont {Sbraccia}}, \bibinfo {author} {\bibfnamefont
  {S.}~\bibnamefont {Scandolo}}, \bibinfo {author} {\bibfnamefont
  {G.}~\bibnamefont {Sclauzero}}, \bibinfo {author} {\bibfnamefont {A.~P.}\
  \bibnamefont {Seitsonen}}, \bibinfo {author} {\bibfnamefont {A.}~\bibnamefont
  {Smogunov}}, \bibinfo {author} {\bibfnamefont {P.}~\bibnamefont {Umari}},\
  and\ \bibinfo {author} {\bibfnamefont {R.~M.}\ \bibnamefont {Wentzcovitch}},\
  }\bibfield  {title} {\bibinfo {title} {{QUANTUM} {ESPRESSO}: a modular and
  open-source software project for quantum simulations of materials},\ }\href
  {https://doi.org/10.1088/0953-8984/21/39/395502} {\bibfield  {journal}
  {\bibinfo  {journal} {J. Phys. Condens. Matter}\ }\textbf {\bibinfo {volume}
  {21}},\ \bibinfo {pages} {395502} (\bibinfo {year} {2009})}\BibitemShut
  {NoStop}%
\bibitem [{\citenamefont {Giannozzi}\ \emph {et~al.}(2017)\citenamefont
  {Giannozzi}, \citenamefont {Andreussi}, \citenamefont {Brumme}, \citenamefont
  {Bunau}, \citenamefont {Nardelli}, \citenamefont {Calandra}, \citenamefont
  {Car}, \citenamefont {Cavazzoni}, \citenamefont {Ceresoli}, \citenamefont
  {Cococcioni}, \citenamefont {Colonna}, \citenamefont {Carnimeo},
  \citenamefont {Corso}, \citenamefont {de~Gironcoli}, \citenamefont {Delugas},
  \citenamefont {DiStasio}, \citenamefont {Ferretti}, \citenamefont {Floris},
  \citenamefont {Fratesi}, \citenamefont {Fugallo}, \citenamefont {Gebauer},
  \citenamefont {Gerstmann}, \citenamefont {Giustino}, \citenamefont {Gorni},
  \citenamefont {Jia}, \citenamefont {Kawamura}, \citenamefont {Ko},
  \citenamefont {Kokalj}, \citenamefont {Kü{\c{c}}ükbenli}, \citenamefont
  {Lazzeri}, \citenamefont {Marsili}, \citenamefont {Marzari}, \citenamefont
  {Mauri}, \citenamefont {Nguyen}, \citenamefont {Nguyen}, \citenamefont {de-la
  Roza}, \citenamefont {Paulatto}, \citenamefont {Ponc{\'{e}}}, \citenamefont
  {Rocca}, \citenamefont {Sabatini}, \citenamefont {Santra}, \citenamefont
  {Schlipf}, \citenamefont {Seitsonen}, \citenamefont {Smogunov}, \citenamefont
  {Timrov}, \citenamefont {Thonhauser}, \citenamefont {Umari}, \citenamefont
  {Vast}, \citenamefont {Wu},\ and\ \citenamefont {Baroni}}]{Giannozzi_2017}%
  \BibitemOpen
  \bibfield  {author} {\bibinfo {author} {\bibfnamefont {P.}~\bibnamefont
  {Giannozzi}}, \bibinfo {author} {\bibfnamefont {O.}~\bibnamefont
  {Andreussi}}, \bibinfo {author} {\bibfnamefont {T.}~\bibnamefont {Brumme}},
  \bibinfo {author} {\bibfnamefont {O.}~\bibnamefont {Bunau}}, \bibinfo
  {author} {\bibfnamefont {M.~B.}\ \bibnamefont {Nardelli}}, \bibinfo {author}
  {\bibfnamefont {M.}~\bibnamefont {Calandra}}, \bibinfo {author}
  {\bibfnamefont {R.}~\bibnamefont {Car}}, \bibinfo {author} {\bibfnamefont
  {C.}~\bibnamefont {Cavazzoni}}, \bibinfo {author} {\bibfnamefont
  {D.}~\bibnamefont {Ceresoli}}, \bibinfo {author} {\bibfnamefont
  {M.}~\bibnamefont {Cococcioni}}, \bibinfo {author} {\bibfnamefont
  {N.}~\bibnamefont {Colonna}}, \bibinfo {author} {\bibfnamefont
  {I.}~\bibnamefont {Carnimeo}}, \bibinfo {author} {\bibfnamefont {A.~D.}\
  \bibnamefont {Corso}}, \bibinfo {author} {\bibfnamefont {S.}~\bibnamefont
  {de~Gironcoli}}, \bibinfo {author} {\bibfnamefont {P.}~\bibnamefont
  {Delugas}}, \bibinfo {author} {\bibfnamefont {R.~A.}\ \bibnamefont
  {DiStasio}}, \bibinfo {author} {\bibfnamefont {A.}~\bibnamefont {Ferretti}},
  \bibinfo {author} {\bibfnamefont {A.}~\bibnamefont {Floris}}, \bibinfo
  {author} {\bibfnamefont {G.}~\bibnamefont {Fratesi}}, \bibinfo {author}
  {\bibfnamefont {G.}~\bibnamefont {Fugallo}}, \bibinfo {author} {\bibfnamefont
  {R.}~\bibnamefont {Gebauer}}, \bibinfo {author} {\bibfnamefont
  {U.}~\bibnamefont {Gerstmann}}, \bibinfo {author} {\bibfnamefont
  {F.}~\bibnamefont {Giustino}}, \bibinfo {author} {\bibfnamefont
  {T.}~\bibnamefont {Gorni}}, \bibinfo {author} {\bibfnamefont
  {J.}~\bibnamefont {Jia}}, \bibinfo {author} {\bibfnamefont {M.}~\bibnamefont
  {Kawamura}}, \bibinfo {author} {\bibfnamefont {H.-Y.}\ \bibnamefont {Ko}},
  \bibinfo {author} {\bibfnamefont {A.}~\bibnamefont {Kokalj}}, \bibinfo
  {author} {\bibfnamefont {E.}~\bibnamefont {Kü{\c{c}}ükbenli}}, \bibinfo
  {author} {\bibfnamefont {M.}~\bibnamefont {Lazzeri}}, \bibinfo {author}
  {\bibfnamefont {M.}~\bibnamefont {Marsili}}, \bibinfo {author} {\bibfnamefont
  {N.}~\bibnamefont {Marzari}}, \bibinfo {author} {\bibfnamefont
  {F.}~\bibnamefont {Mauri}}, \bibinfo {author} {\bibfnamefont {N.~L.}\
  \bibnamefont {Nguyen}}, \bibinfo {author} {\bibfnamefont {H.-V.}\
  \bibnamefont {Nguyen}}, \bibinfo {author} {\bibfnamefont {A.~O.}\
  \bibnamefont {de-la Roza}}, \bibinfo {author} {\bibfnamefont
  {L.}~\bibnamefont {Paulatto}}, \bibinfo {author} {\bibfnamefont
  {S.}~\bibnamefont {Ponc{\'{e}}}}, \bibinfo {author} {\bibfnamefont
  {D.}~\bibnamefont {Rocca}}, \bibinfo {author} {\bibfnamefont
  {R.}~\bibnamefont {Sabatini}}, \bibinfo {author} {\bibfnamefont
  {B.}~\bibnamefont {Santra}}, \bibinfo {author} {\bibfnamefont
  {M.}~\bibnamefont {Schlipf}}, \bibinfo {author} {\bibfnamefont {A.~P.}\
  \bibnamefont {Seitsonen}}, \bibinfo {author} {\bibfnamefont {A.}~\bibnamefont
  {Smogunov}}, \bibinfo {author} {\bibfnamefont {I.}~\bibnamefont {Timrov}},
  \bibinfo {author} {\bibfnamefont {T.}~\bibnamefont {Thonhauser}}, \bibinfo
  {author} {\bibfnamefont {P.}~\bibnamefont {Umari}}, \bibinfo {author}
  {\bibfnamefont {N.}~\bibnamefont {Vast}}, \bibinfo {author} {\bibfnamefont
  {X.}~\bibnamefont {Wu}},\ and\ \bibinfo {author} {\bibfnamefont
  {S.}~\bibnamefont {Baroni}},\ }\bibfield  {title} {\bibinfo {title} {Advanced
  capabilities for materials modelling with {Quantum} {ESPRESSO}},\ }\href
  {https://doi.org/10.1088/1361-648x/aa8f79} {\bibfield  {journal} {\bibinfo
  {journal} {J. Phys. Condens. Matter}\ }\textbf {\bibinfo {volume} {29}},\
  \bibinfo {pages} {465901} (\bibinfo {year} {2017})}\BibitemShut {NoStop}%
\bibitem [{psl()}]{pslibrary}%
  \BibitemOpen
  \href@noop {} {}\bibinfo {howpublished}
  {https://dalcorso.github.io/pslibrary/}\BibitemShut {NoStop}%
\bibitem [{\citenamefont {{Dal Corso}}(2014)}]{DALCORSO2014337}%
  \BibitemOpen
  \bibfield  {author} {\bibinfo {author} {\bibfnamefont {A.}~\bibnamefont {{Dal
  Corso}}},\ }\bibfield  {title} {\bibinfo {title} {Pseudopotentials periodic
  table: From {H} to {Pu}},\ }\href
  {https://doi.org/https://doi.org/10.1016/j.commatsci.2014.07.043} {\bibfield
  {journal} {\bibinfo  {journal} {Comput. Mater. Sci.}\ }\textbf {\bibinfo
  {volume} {95}},\ \bibinfo {pages} {337} (\bibinfo {year} {2014})}\BibitemShut
  {NoStop}%
\bibitem [{ope()}]{openmx}%
  \BibitemOpen
  \href@noop {} {}\bibinfo {howpublished}
  {http://www.openmx-square.org}\BibitemShut {NoStop}%
\bibitem [{\citenamefont {Ozaki}(2003)}]{PhysRevB.67.155108}%
  \BibitemOpen
  \bibfield  {author} {\bibinfo {author} {\bibfnamefont {T.}~\bibnamefont
  {Ozaki}},\ }\bibfield  {title} {\bibinfo {title} {Variationally optimized
  atomic orbitals for large-scale electronic structures},\ }\href
  {https://doi.org/10.1103/PhysRevB.67.155108} {\bibfield  {journal} {\bibinfo
  {journal} {Phys. Rev. B}\ }\textbf {\bibinfo {volume} {67}},\ \bibinfo
  {pages} {155108} (\bibinfo {year} {2003})}\BibitemShut {NoStop}%
\bibitem [{\citenamefont {Ozaki}\ and\ \citenamefont
  {Kino}(2004)}]{PhysRevB.69.195113}%
  \BibitemOpen
  \bibfield  {author} {\bibinfo {author} {\bibfnamefont {T.}~\bibnamefont
  {Ozaki}}\ and\ \bibinfo {author} {\bibfnamefont {H.}~\bibnamefont {Kino}},\
  }\bibfield  {title} {\bibinfo {title} {Numerical atomic basis orbitals from
  {H} to {Kr}},\ }\href {https://doi.org/10.1103/PhysRevB.69.195113} {\bibfield
   {journal} {\bibinfo  {journal} {Phys. Rev. B}\ }\textbf {\bibinfo {volume}
  {69}},\ \bibinfo {pages} {195113} (\bibinfo {year} {2004})}\BibitemShut
  {NoStop}%
\bibitem [{\citenamefont {Ozaki}\ and\ \citenamefont
  {Kino}(2005)}]{PhysRevB.72.045121}%
  \BibitemOpen
  \bibfield  {author} {\bibinfo {author} {\bibfnamefont {T.}~\bibnamefont
  {Ozaki}}\ and\ \bibinfo {author} {\bibfnamefont {H.}~\bibnamefont {Kino}},\
  }\bibfield  {title} {\bibinfo {title} {Efficient projector expansion for the
  ab initio {LCAO} method},\ }\href
  {https://doi.org/10.1103/PhysRevB.72.045121} {\bibfield  {journal} {\bibinfo
  {journal} {Phys. Rev. B}\ }\textbf {\bibinfo {volume} {72}},\ \bibinfo
  {pages} {045121} (\bibinfo {year} {2005})}\BibitemShut {NoStop}%
\bibitem [{\citenamefont {Thonhauser}\ \emph {et~al.}(2007)\citenamefont
  {Thonhauser}, \citenamefont {Cooper}, \citenamefont {Li}, \citenamefont
  {Puzder}, \citenamefont {Hyldgaard},\ and\ \citenamefont
  {Langreth}}]{PhysRevB.76.125112}%
  \BibitemOpen
  \bibfield  {author} {\bibinfo {author} {\bibfnamefont {T.}~\bibnamefont
  {Thonhauser}}, \bibinfo {author} {\bibfnamefont {V.~R.}\ \bibnamefont
  {Cooper}}, \bibinfo {author} {\bibfnamefont {S.}~\bibnamefont {Li}}, \bibinfo
  {author} {\bibfnamefont {A.}~\bibnamefont {Puzder}}, \bibinfo {author}
  {\bibfnamefont {P.}~\bibnamefont {Hyldgaard}},\ and\ \bibinfo {author}
  {\bibfnamefont {D.~C.}\ \bibnamefont {Langreth}},\ }\bibfield  {title}
  {\bibinfo {title} {Van der {Waals} density functional: Self-consistent
  potential and the nature of the van der {Waals} bond},\ }\href
  {https://doi.org/10.1103/PhysRevB.76.125112} {\bibfield  {journal} {\bibinfo
  {journal} {Phys. Rev. B}\ }\textbf {\bibinfo {volume} {76}},\ \bibinfo
  {pages} {125112} (\bibinfo {year} {2007})}\BibitemShut {NoStop}%
\bibitem [{\citenamefont {Hamada}(2014)}]{PhysRevB.89.121103}%
  \BibitemOpen
  \bibfield  {author} {\bibinfo {author} {\bibfnamefont {I.}~\bibnamefont
  {Hamada}},\ }\bibfield  {title} {\bibinfo {title} {van der {Waals} density
  functional made accurate},\ }\href
  {https://doi.org/10.1103/PhysRevB.89.121103} {\bibfield  {journal} {\bibinfo
  {journal} {Phys. Rev. B}\ }\textbf {\bibinfo {volume} {89}},\ \bibinfo
  {pages} {121103} (\bibinfo {year} {2014})}\BibitemShut {NoStop}%
\bibitem [{\citenamefont {Perdew}\ \emph {et~al.}(1996)\citenamefont {Perdew},
  \citenamefont {Burke},\ and\ \citenamefont
  {Ernzerhof}}]{PhysRevLett.77.3865}%
  \BibitemOpen
  \bibfield  {author} {\bibinfo {author} {\bibfnamefont {J.~P.}\ \bibnamefont
  {Perdew}}, \bibinfo {author} {\bibfnamefont {K.}~\bibnamefont {Burke}},\ and\
  \bibinfo {author} {\bibfnamefont {M.}~\bibnamefont {Ernzerhof}},\ }\bibfield
  {title} {\bibinfo {title} {Generalized gradient approximation made simple},\
  }\href {https://doi.org/10.1103/PhysRevLett.77.3865} {\bibfield  {journal}
  {\bibinfo  {journal} {Phys. Rev. Lett.}\ }\textbf {\bibinfo {volume} {77}},\
  \bibinfo {pages} {3865} (\bibinfo {year} {1996})}\BibitemShut {NoStop}%
\bibitem [{\citenamefont {Tomanek}\ \emph {et~al.}(1988)\citenamefont
  {Tomanek}, \citenamefont {Wentzcovitch}, \citenamefont {Louie},\ and\
  \citenamefont {Cohen}}]{PhysRevB.37.3134}%
  \BibitemOpen
  \bibfield  {author} {\bibinfo {author} {\bibfnamefont {D.}~\bibnamefont
  {Tomanek}}, \bibinfo {author} {\bibfnamefont {R.~M.}\ \bibnamefont
  {Wentzcovitch}}, \bibinfo {author} {\bibfnamefont {S.~G.}\ \bibnamefont
  {Louie}},\ and\ \bibinfo {author} {\bibfnamefont {M.~L.}\ \bibnamefont
  {Cohen}},\ }\bibfield  {title} {\bibinfo {title} {Calculation of electronic
  and structural properties of {${\mathrm{BC}}_{3}$}},\ }\href
  {https://doi.org/10.1103/PhysRevB.37.3134} {\bibfield  {journal} {\bibinfo
  {journal} {Phys. Rev. B}\ }\textbf {\bibinfo {volume} {37}},\ \bibinfo
  {pages} {3134} (\bibinfo {year} {1988})}\BibitemShut {NoStop}%
\bibitem [{\citenamefont {Lin}\ and\ \citenamefont
  {Ni}(2012)}]{doi:10.1063/1.3681899}%
  \BibitemOpen
  \bibfield  {author} {\bibinfo {author} {\bibfnamefont {X.}~\bibnamefont
  {Lin}}\ and\ \bibinfo {author} {\bibfnamefont {J.}~\bibnamefont {Ni}},\
  }\bibfield  {title} {\bibinfo {title} {Electronic and magnetic properties of
  substitutionally {Fe}-, {Co}-, and {Ni}-doped {BC$_3$} honeycomb structure},\
  }\href {https://doi.org/10.1063/1.3681899} {\bibfield  {journal} {\bibinfo
  {journal} {J. Appl. Phys.}\ }\textbf {\bibinfo {volume} {111}},\ \bibinfo
  {pages} {034309} (\bibinfo {year} {2012})}\BibitemShut {NoStop}%
\bibitem [{\citenamefont {Tan}\ \emph {et~al.}(2014)\citenamefont {Tan},
  \citenamefont {Jin},\ and\ \citenamefont {Chen}}]{C3CP54838D}%
  \BibitemOpen
  \bibfield  {author} {\bibinfo {author} {\bibfnamefont {X.}~\bibnamefont
  {Tan}}, \bibinfo {author} {\bibfnamefont {P.}~\bibnamefont {Jin}},\ and\
  \bibinfo {author} {\bibfnamefont {Z.}~\bibnamefont {Chen}},\ }\bibfield
  {title} {\bibinfo {title} {With the same clar formulas{,} do the
  two-dimensional sandwich nanostructures {X–Cr–X} ({X} = {C$_4$H}{,}
  {NC$_3$} and {BC$_3$}) behave similarly?},\ }\href
  {https://doi.org/10.1039/C3CP54838D} {\bibfield  {journal} {\bibinfo
  {journal} {Phys. Chem. Chem. Phys.}\ }\textbf {\bibinfo {volume} {16}},\
  \bibinfo {pages} {6002} (\bibinfo {year} {2014})}\BibitemShut {NoStop}%
\bibitem [{\citenamefont {Pizzi}\ \emph {et~al.}(2020)\citenamefont {Pizzi},
  \citenamefont {Vitale}, \citenamefont {Arita}, \citenamefont {Blügel},
  \citenamefont {Freimuth}, \citenamefont {G{\'{e}}ranton}, \citenamefont
  {Gibertini}, \citenamefont {Gresch}, \citenamefont {Johnson}, \citenamefont
  {Koretsune}, \citenamefont {Iba{\~{n}}ez-Azpiroz}, \citenamefont {Lee},
  \citenamefont {Lihm}, \citenamefont {Marchand}, \citenamefont {Marrazzo},
  \citenamefont {Mokrousov}, \citenamefont {Mustafa}, \citenamefont {Nohara},
  \citenamefont {Nomura}, \citenamefont {Paulatto}, \citenamefont
  {Ponc{\'{e}}}, \citenamefont {Ponweiser}, \citenamefont {Qiao}, \citenamefont
  {Thöle}, \citenamefont {Tsirkin}, \citenamefont {Wierzbowska}, \citenamefont
  {Marzari}, \citenamefont {Vanderbilt}, \citenamefont {Souza}, \citenamefont
  {Mostofi},\ and\ \citenamefont {Yates}}]{Pizzi2020}%
  \BibitemOpen
  \bibfield  {author} {\bibinfo {author} {\bibfnamefont {G.}~\bibnamefont
  {Pizzi}}, \bibinfo {author} {\bibfnamefont {V.}~\bibnamefont {Vitale}},
  \bibinfo {author} {\bibfnamefont {R.}~\bibnamefont {Arita}}, \bibinfo
  {author} {\bibfnamefont {S.}~\bibnamefont {Blügel}}, \bibinfo {author}
  {\bibfnamefont {F.}~\bibnamefont {Freimuth}}, \bibinfo {author}
  {\bibfnamefont {G.}~\bibnamefont {G{\'{e}}ranton}}, \bibinfo {author}
  {\bibfnamefont {M.}~\bibnamefont {Gibertini}}, \bibinfo {author}
  {\bibfnamefont {D.}~\bibnamefont {Gresch}}, \bibinfo {author} {\bibfnamefont
  {C.}~\bibnamefont {Johnson}}, \bibinfo {author} {\bibfnamefont
  {T.}~\bibnamefont {Koretsune}}, \bibinfo {author} {\bibfnamefont
  {J.}~\bibnamefont {Iba{\~{n}}ez-Azpiroz}}, \bibinfo {author} {\bibfnamefont
  {H.}~\bibnamefont {Lee}}, \bibinfo {author} {\bibfnamefont {J.-M.}\
  \bibnamefont {Lihm}}, \bibinfo {author} {\bibfnamefont {D.}~\bibnamefont
  {Marchand}}, \bibinfo {author} {\bibfnamefont {A.}~\bibnamefont {Marrazzo}},
  \bibinfo {author} {\bibfnamefont {Y.}~\bibnamefont {Mokrousov}}, \bibinfo
  {author} {\bibfnamefont {J.~I.}\ \bibnamefont {Mustafa}}, \bibinfo {author}
  {\bibfnamefont {Y.}~\bibnamefont {Nohara}}, \bibinfo {author} {\bibfnamefont
  {Y.}~\bibnamefont {Nomura}}, \bibinfo {author} {\bibfnamefont
  {L.}~\bibnamefont {Paulatto}}, \bibinfo {author} {\bibfnamefont
  {S.}~\bibnamefont {Ponc{\'{e}}}}, \bibinfo {author} {\bibfnamefont
  {T.}~\bibnamefont {Ponweiser}}, \bibinfo {author} {\bibfnamefont
  {J.}~\bibnamefont {Qiao}}, \bibinfo {author} {\bibfnamefont {F.}~\bibnamefont
  {Thöle}}, \bibinfo {author} {\bibfnamefont {S.~S.}\ \bibnamefont {Tsirkin}},
  \bibinfo {author} {\bibfnamefont {M.}~\bibnamefont {Wierzbowska}}, \bibinfo
  {author} {\bibfnamefont {N.}~\bibnamefont {Marzari}}, \bibinfo {author}
  {\bibfnamefont {D.}~\bibnamefont {Vanderbilt}}, \bibinfo {author}
  {\bibfnamefont {I.}~\bibnamefont {Souza}}, \bibinfo {author} {\bibfnamefont
  {A.~A.}\ \bibnamefont {Mostofi}},\ and\ \bibinfo {author} {\bibfnamefont
  {J.~R.}\ \bibnamefont {Yates}},\ }\bibfield  {title} {\bibinfo {title}
  {Wannier90 as a community code: new features and applications},\ }\href
  {https://doi.org/10.1088/1361-648x/ab51ff} {\bibfield  {journal} {\bibinfo
  {journal} {J. Phys. Condens. Matter}\ }\textbf {\bibinfo {volume} {32}},\
  \bibinfo {pages} {165902} (\bibinfo {year} {2020})}\BibitemShut {NoStop}%
\bibitem [{\citenamefont {Hsing}\ \emph {et~al.}(2014)\citenamefont {Hsing},
  \citenamefont {Cheng}, \citenamefont {Chou}, \citenamefont {Chang},\ and\
  \citenamefont {Wei}}]{Hsing_2014}%
  \BibitemOpen
  \bibfield  {author} {\bibinfo {author} {\bibfnamefont {C.-R.}\ \bibnamefont
  {Hsing}}, \bibinfo {author} {\bibfnamefont {C.}~\bibnamefont {Cheng}},
  \bibinfo {author} {\bibfnamefont {J.-P.}\ \bibnamefont {Chou}}, \bibinfo
  {author} {\bibfnamefont {C.-M.}\ \bibnamefont {Chang}},\ and\ \bibinfo
  {author} {\bibfnamefont {C.-M.}\ \bibnamefont {Wei}},\ }\bibfield  {title}
  {\bibinfo {title} {Van der {Waals} interaction in a boron nitride bilayer},\
  }\href {https://doi.org/10.1088/1367-2630/16/11/113015} {\bibfield  {journal}
  {\bibinfo  {journal} {New J. Phys.}\ }\textbf {\bibinfo {volume} {16}},\
  \bibinfo {pages} {113015} (\bibinfo {year} {2014})}\BibitemShut {NoStop}%
\bibitem [{\citenamefont {Marzari}\ and\ \citenamefont
  {Vanderbilt}(1997)}]{PhysRevB.56.12847}%
  \BibitemOpen
  \bibfield  {author} {\bibinfo {author} {\bibfnamefont {N.}~\bibnamefont
  {Marzari}}\ and\ \bibinfo {author} {\bibfnamefont {D.}~\bibnamefont
  {Vanderbilt}},\ }\bibfield  {title} {\bibinfo {title} {Maximally localized
  generalized {Wannier} functions for composite energy bands},\ }\href
  {https://doi.org/10.1103/PhysRevB.56.12847} {\bibfield  {journal} {\bibinfo
  {journal} {Phys. Rev. B}\ }\textbf {\bibinfo {volume} {56}},\ \bibinfo
  {pages} {12847} (\bibinfo {year} {1997})}\BibitemShut {NoStop}%
\bibitem [{\citenamefont {Kugel}\ and\ \citenamefont
  {Khomski{\u{\i}}}(1982)}]{Kugel1982}%
  \BibitemOpen
  \bibfield  {author} {\bibinfo {author} {\bibfnamefont {K.~I.}\ \bibnamefont
  {Kugel}}\ and\ \bibinfo {author} {\bibfnamefont {D.~I.}\ \bibnamefont
  {Khomski{\u{\i}}}},\ }\bibfield  {title} {\bibinfo {title} {The
  {Jahn}-{Teller} effect and magnetism: transition metal compounds},\ }\href
  {https://doi.org/10.1070/pu1982v025n04abeh004537} {\bibfield  {journal}
  {\bibinfo  {journal} {Soviet Physics Uspekhi}\ }\textbf {\bibinfo {volume}
  {25}},\ \bibinfo {pages} {231} (\bibinfo {year} {1982})}\BibitemShut
  {NoStop}%
\bibitem [{\citenamefont {Khaliullin}\ and\ \citenamefont
  {Oudovenko}(1997)}]{PhysRevB.56.R14243}%
  \BibitemOpen
  \bibfield  {author} {\bibinfo {author} {\bibfnamefont {G.}~\bibnamefont
  {Khaliullin}}\ and\ \bibinfo {author} {\bibfnamefont {V.}~\bibnamefont
  {Oudovenko}},\ }\bibfield  {title} {\bibinfo {title} {Spin and orbital
  excitation spectrum in the {Kugel}-{Khomskii} model},\ }\href
  {https://doi.org/10.1103/PhysRevB.56.R14243} {\bibfield  {journal} {\bibinfo
  {journal} {Phys. Rev. B}\ }\textbf {\bibinfo {volume} {56}},\ \bibinfo
  {pages} {R14243} (\bibinfo {year} {1997})}\BibitemShut {NoStop}%
\bibitem [{\citenamefont {Wang}\ and\ \citenamefont
  {Vishwanath}(2009)}]{PhysRevB.80.064413}%
  \BibitemOpen
  \bibfield  {author} {\bibinfo {author} {\bibfnamefont {F.}~\bibnamefont
  {Wang}}\ and\ \bibinfo {author} {\bibfnamefont {A.}~\bibnamefont
  {Vishwanath}},\ }\bibfield  {title} {\bibinfo {title} {{${\text{Z}}_{2}$}
  spin-orbital liquid state in the square lattice {Kugel}-{Khomskii} model},\
  }\href {https://doi.org/10.1103/PhysRevB.80.064413} {\bibfield  {journal}
  {\bibinfo  {journal} {Phys. Rev. B}\ }\textbf {\bibinfo {volume} {80}},\
  \bibinfo {pages} {064413} (\bibinfo {year} {2009})}\BibitemShut {NoStop}%
\bibitem [{\citenamefont {Giamarchi}(2003)}]{Giamarchi_1D}%
  \BibitemOpen
  \bibfield  {author} {\bibinfo {author} {\bibfnamefont {T.}~\bibnamefont
  {Giamarchi}},\ }\href@noop {} {\emph {\bibinfo {title} {Quantum Physics in
  One Dimension}}}\ (\bibinfo  {publisher} {Clarendon Press},\ \bibinfo
  {address} {Oxford},\ \bibinfo {year} {2003})\BibitemShut {NoStop}%
\bibitem [{\citenamefont {Cea}\ \emph {et~al.}(2019)\citenamefont {Cea},
  \citenamefont {Walet},\ and\ \citenamefont {Guinea}}]{PhysRevB.100.205113}%
  \BibitemOpen
  \bibfield  {author} {\bibinfo {author} {\bibfnamefont {T.}~\bibnamefont
  {Cea}}, \bibinfo {author} {\bibfnamefont {N.~R.}\ \bibnamefont {Walet}},\
  and\ \bibinfo {author} {\bibfnamefont {F.}~\bibnamefont {Guinea}},\
  }\bibfield  {title} {\bibinfo {title} {Electronic band structure and pinning
  of {Fermi} energy to {Van Hove} singularities in twisted bilayer graphene: A
  self-consistent approach},\ }\href
  {https://doi.org/10.1103/PhysRevB.100.205113} {\bibfield  {journal} {\bibinfo
   {journal} {Phys. Rev. B}\ }\textbf {\bibinfo {volume} {100}},\ \bibinfo
  {pages} {205113} (\bibinfo {year} {2019})}\BibitemShut {NoStop}%
\bibitem [{\citenamefont {Kang}\ and\ \citenamefont
  {Vafek}(2020)}]{PhysRevB.102.035161}%
  \BibitemOpen
  \bibfield  {author} {\bibinfo {author} {\bibfnamefont {J.}~\bibnamefont
  {Kang}}\ and\ \bibinfo {author} {\bibfnamefont {O.}~\bibnamefont {Vafek}},\
  }\bibfield  {title} {\bibinfo {title} {Non-{Abelian} {Dirac} node braiding
  and near-degeneracy of correlated phases at odd integer filling in
  magic-angle twisted bilayer graphene},\ }\href
  {https://doi.org/10.1103/PhysRevB.102.035161} {\bibfield  {journal} {\bibinfo
   {journal} {Phys. Rev. B}\ }\textbf {\bibinfo {volume} {102}},\ \bibinfo
  {pages} {035161} (\bibinfo {year} {2020})}\BibitemShut {NoStop}%
\bibitem [{\citenamefont {Cea}\ and\ \citenamefont
  {Guinea}(2020)}]{PhysRevB.102.045107}%
  \BibitemOpen
  \bibfield  {author} {\bibinfo {author} {\bibfnamefont {T.}~\bibnamefont
  {Cea}}\ and\ \bibinfo {author} {\bibfnamefont {F.}~\bibnamefont {Guinea}},\
  }\bibfield  {title} {\bibinfo {title} {Band structure and insulating states
  driven by {Coulomb} interaction in twisted bilayer graphene},\ }\href
  {https://doi.org/10.1103/PhysRevB.102.045107} {\bibfield  {journal} {\bibinfo
   {journal} {Phys. Rev. B}\ }\textbf {\bibinfo {volume} {102}},\ \bibinfo
  {pages} {045107} (\bibinfo {year} {2020})}\BibitemShut {NoStop}%
\bibitem [{\citenamefont {Wang}\ \emph {et~al.}(2022)\citenamefont {Wang},
  \citenamefont {Yu}, \citenamefont {Kwan}, \citenamefont {Jia}, \citenamefont
  {Lei}, \citenamefont {Klemenz}, \citenamefont {Cevallos}, \citenamefont
  {Singha}, \citenamefont {Devakul}, \citenamefont {Watanabe}, \citenamefont
  {Taniguchi}, \citenamefont {Sondhi}, \citenamefont {Cava}, \citenamefont
  {Schoop}, \citenamefont {Parameswaran},\ and\ \citenamefont
  {Wu}}]{wang2022-luttinger}%
  \BibitemOpen
  \bibfield  {author} {\bibinfo {author} {\bibfnamefont {P.}~\bibnamefont
  {Wang}}, \bibinfo {author} {\bibfnamefont {G.}~\bibnamefont {Yu}}, \bibinfo
  {author} {\bibfnamefont {Y.~H.}\ \bibnamefont {Kwan}}, \bibinfo {author}
  {\bibfnamefont {Y.}~\bibnamefont {Jia}}, \bibinfo {author} {\bibfnamefont
  {S.}~\bibnamefont {Lei}}, \bibinfo {author} {\bibfnamefont {S.}~\bibnamefont
  {Klemenz}}, \bibinfo {author} {\bibfnamefont {F.~A.}\ \bibnamefont
  {Cevallos}}, \bibinfo {author} {\bibfnamefont {R.}~\bibnamefont {Singha}},
  \bibinfo {author} {\bibfnamefont {T.}~\bibnamefont {Devakul}}, \bibinfo
  {author} {\bibfnamefont {K.}~\bibnamefont {Watanabe}}, \bibinfo {author}
  {\bibfnamefont {T.}~\bibnamefont {Taniguchi}}, \bibinfo {author}
  {\bibfnamefont {S.~L.}\ \bibnamefont {Sondhi}}, \bibinfo {author}
  {\bibfnamefont {R.~J.}\ \bibnamefont {Cava}}, \bibinfo {author}
  {\bibfnamefont {L.~M.}\ \bibnamefont {Schoop}}, \bibinfo {author}
  {\bibfnamefont {S.~A.}\ \bibnamefont {Parameswaran}},\ and\ \bibinfo {author}
  {\bibfnamefont {S.}~\bibnamefont {Wu}},\ }\bibfield  {title} {\bibinfo
  {title} {One-dimensional {Luttinger} liquids in a two-dimensional moir{\'e}
  lattice},\ }\href {https://doi.org/10.1038/s41586-022-04514-6} {\bibfield
  {journal} {\bibinfo  {journal} {Nature}\ }\textbf {\bibinfo {volume} {605}},\
  \bibinfo {pages} {57} (\bibinfo {year} {2022})}\BibitemShut {NoStop}%
\bibitem [{\citenamefont {Kennes}\ \emph {et~al.}(2020)\citenamefont {Kennes},
  \citenamefont {Xian}, \citenamefont {Claassen},\ and\ \citenamefont
  {Rubio}}]{kennes2020one-b}%
  \BibitemOpen
  \bibfield  {author} {\bibinfo {author} {\bibfnamefont {D.~M.}\ \bibnamefont
  {Kennes}}, \bibinfo {author} {\bibfnamefont {L.}~\bibnamefont {Xian}},
  \bibinfo {author} {\bibfnamefont {M.}~\bibnamefont {Claassen}},\ and\
  \bibinfo {author} {\bibfnamefont {A.}~\bibnamefont {Rubio}},\ }\bibfield
  {title} {\bibinfo {title} {One-dimensional flat bands in twisted bilayer
  germanium selenide},\ }\href {https://doi.org/10.1038/s41467-020-14947-0}
  {\bibfield  {journal} {\bibinfo  {journal} {Nat. Commun.}\ }\textbf {\bibinfo
  {volume} {11}},\ \bibinfo {pages} {1124} (\bibinfo {year}
  {2020})}\BibitemShut {NoStop}%
\end{thebibliography}
\end{document}